\documentclass{article}
\usepackage[colorlinks]{hyperref}
\usepackage[italian,english]{babel}
\usepackage[textheight=24 cm, textwidth=18cm]{geometry}
\usepackage[page,toc,titletoc,title]{appendix}
\usepackage{graphicx}
\begin{document}
  \hypersetup{
 colorlinks   = true, 
  urlcolor     = blue, 
  linkcolor    = blue, 
  citecolor   = blue 
}
\renewcommand\bibname{References}
\thispagestyle{empty}
\begin{center}
{\Large\bf Teaching special relativity in elementary
physics or upper high school courses}
\end{center}
\vskip4mm {\bf Maria Grazia Blumetti, Biagio Buonaura, Giuseppe Giuliani and Marco Litterio}
\vskip4mm\par\noindent
{\small Liceo Scientifico Galilei, Potenza -- Formerly, ISIS Albertini, Via Circumvallazione
292, 80035 Nola, Italy -- Formerly,  Dipartimento di Fisica, Universit\`a
degli Studi di Pavia, via Bassi 6, 27100 Pavia, Italy}  -- Liceo Scientifico Labriola, Roma
\vskip4mm\par\noindent
{\verb"marziablu@yahoo.it"}, {\verb"bbuonaura@gmail.com"}, {\verb"giuseppe.giuliani@unipv.it"}, {\verb"marcolitprof@gmail.com"}
\vskip4mm\par\noindent
{\bf Abstract.} {This paper aims to provide teachers with a tool to teach the essential features of special relativity, considering the students' difficulties highlighted by numerous studies. Our proposal presents special relativity as the solution to the troubles of Newtonian dynamics, exemplified by the infinities of Newtonian uniformly accelerated motion. The paper's main section uses thought experiments with the exchange of flashes of light of null duration between two inertial reference frames to derive the kinematics effect of special relativity (time dilation, length contraction, Doppler effect, relativity of simultaneity, and Lorentz transformations). Simulations illustrate the results of the simple calculations. The discussion of experimental corroborations of the kinematics effects of special relativity complements the theoretical treatments. The Doppler effect, typically treated within the wave description of light, is addressed as an application of energy and linear momentum conservation during the emission or absorption of a photon by an atom (or a nucleus). When opportune, the paper suggests implementing teaching practices in the topics developed for teachers. Two of us' preliminary tests in the classroom ask for a wider one, including standard evaluation procedures of students' learning.}
\tableofcontents

\section{Introduction}
At the beginning of the twentieth century, Physics underwent a dramatic transformation. The previously held continuous view of matter and radiation gave way to a discrete understanding, and Einstein's theory of special relativity compelled scientists to move beyond Newton's concepts of space and time. The hypothesis of a maximum speed -- the speed of light in a vacuum -- necessitated the development of new electrodynamics and dynamics.
\par
In the nineteenth century, some physicists -- notably Ludwig Boltzmann -- considered atoms possible constituents of matter. However, atoms were considered devoid of any internal structure. After Joseph John Thomson discovered the electron (1897), the internal structure of atoms gradually became the object of theoretical and experimental studies.
The passage from a continuous conception of electromagnetic radiation -- described in terms of electromagnetic waves -- to the idea that light can also be described in terms of light quanta (Einstein, 1905) was slow and challenged until the mid-twenties \cite{ein05lq}.
\par
In physics education, the debate of {\em when} and {\em how} to introduce students to the new physics has a long tradition \cite{review}. In a paper provocatively titled ``What happened to modern Physics?'', Paul Shabajee and Keith Postlethwaite wrote in the abstract \cite{what}: ``Relativity, Quantum Mechanics and Chaos theory are three of the most significant
scientific advances of the 20th Century - each fundamentally changing our
understanding of the physical universe. [\dots] Children and young people
regularly come into contact with the language, concepts, and implications of these
theories through the media and new technologies, and they are the basis of
many contemporary scientific and technological developments. There is surely,
therefore, an urgent need to include the concepts of `20th Century physics' within the
curriculum.''
\par
We shall limit our discussion to introducing the postulates and the implications of special relativity. Several issues arise from the literature. For a compendium of students' conceptual difficulties, see the valuable review \cite{review}. We shall pick only a few suitable to begin our discourse.
\begin{enumerate}
\item  When students encounter the conceptual framework of special relativity, they draw on their existing knowledge developed through daily experiences and their previous acquaintance with Galilean relativity and Newtonian dynamics. At this point, it can be challenging for them to let go of their established views in favor of Einstein's innovative concepts. Therefore, some have suggested introducing the principles of special relativity at an earlier age to better facilitate this transition(14-15 years)\cite{earlier}.
\item Frames of reference: inertial and accelerated. Many students struggle to understand the concept of a frame of reference and often find it challenging to distinguish between these two categories. These difficulties typically arise early in their studies of Galilean relativity and Newtonian dynamics. A study conducted on a group of Physics Education or Science Education students revealed that they tend to view frames of reference as objects with limited dimensions, disregarding events that occur beyond this confined space \cite{baltici}. While students may intuitively differentiate between inertial and non-inertial frames of reference, they often fail to use a clear and consistent criterion for making these distinctions. \cite{india}.
\item  The speed of light in a vacuum is a universal limit, remaining constant across all inertial reference frames. Students often struggle to embrace this fundamental principle because it contradicts their everyday experiences and the Galilean rules of velocity addition. Nonetheless, further studies are needed in this area. Research documented in \cite{luce1} and \cite{luce2} has primarily focused on pre-instructional beliefs regarding the speed of light, with a notable emphasis on the misconception that the ``true'' speed of light can be measured in the reference frame of the light source.
\item Time dilation and length contraction conceived as a non-reciprocal effect, i.e., occurring only in the ``moving'' reference frame \cite{timedilation1, timedilation2}.\label{TD}
\item \label{signal} An in-depth study of undergraduate and graduate student's understanding of the relativity of simultaneity surfaced several rooted beliefs difficult to remove \cite{simul}. The main findings were: ``After instruction, more
than two-thirds of physics undergraduates and one-third of
graduate students in physics are unable to apply the construct
of a reference frame in determining whether or not two
events are simultaneous. Many students interpret the phrase
`relativity of simultaneity' as implying that the simultaneity
of events is determined by an observer on the basis of the
{\em reception} of light signals. They often attribute the relativity
of simultaneity to the difference in signal travel time for
different observers. In this way, they reconcile statements of
the relativity of simultaneity with a belief in absolute simultaneity
and fail to confront the startling ideas of special relativity'' \cite[p. S34]{simul}.
\end{enumerate}
This situation challenges our teaching approaches and tools. A variety of proposals have been tested for their effectiveness using standard methods \cite{review}. These proposals utilize thought experiments, simulations, and more recently, multimedia tools. Another approach emphasizes the importance of a historical and epistemological perspective.

Typically, the physics content to be taught is viewed as fixed, as if it were permanently established in textbooks and accepted teaching practices. Consequently, a critical reassessment of some fundamental concepts is often deemed unnecessary.
Instead, at least two pivotal issues ask for a conceptual and epistemological deepening:
\begin{itemize}
\item[$\diamond$] The concept of `clock' and its role in physical theories and experiments.
\item [$\diamond$] The implications of the fact that light is a relativistic phenomenon because its speed is a speed limit and because the emission or absorption of a photon by an atom or a nucleus must be treated with relativistic dynamics, thus obtaining the Doppler effect within a corpuscular description of light.
\end{itemize}
Accordingly, this paper proposes an approach to teaching special relativity in elementary physics courses or in high school, focusing on the following points\footnote{With the term ``elementary physics courses,'' we refer to courses whose level is intermediate between high school and university. As for high schools, the students' mathematical and physical background knowledge varies widely from country to country. As for Italy, we refer to the last three years of scientific Lyceum.}:
\begin{enumerate}
\item An analysis of what a clock is.\label{clock}
\item An operative definition of inertial and accelerated reference frames.\label{opdef}
\item The necessity of a limit speed emerging from a critical analysis of the accelerated motion due to a constant force in Newtonian mechanics.\label{motoacc}
\item The introduction and discussion of the kinematics relativistic effects using the exchange of light signals of ideally null duration between two inertial frames. This approach also allows the simple introduction of the Doppler effect for light.\label{flashes}
\item The opportunity to introduce some basic formulae of relativistic dynamics without proof. Then, it will be possible to deal with the emission or absorption of a photon by an atom or a nucleus, thus obtaining the formulae of the Doppler effect for photons and discussing their relation with the corresponding formula for light waves. Moreover, the treatment of the absorption of a photon by an atom within Newtonian dynamics leads to the emergence of $mc^2$, the rest energy of a mass.
\item The necessity of discussing the relation between the wave and the corpuscular description of light: the wave description of light yields the correct predictions when the number of photons involved is statistically significant.
\end{enumerate}
The following sections are designed specifically for teachers. Within each section, we provide suggestions for integrating the discussed topics into teaching practices. We explore how these implementations can assist students in moving beyond their spontaneous or entrenched views, thereby alleviating their learning challenges. \textsf{These portions are presented in a distinct font to enhance clarity and facilitate reader comprehension}.

In many teaching contexts, only a subset of the issues mentioned can be effectively addressed. Teachers should choose their approach based on the prior knowledge of their students and their own personal preferences. The material presented in this paper requires teachers to make a dedicated effort to tailor it to their specific teaching environment.
\par
This paper is structured as follows. Section \ref{clocksec} focuses on clocks and their role in physical theories and experiments. A key point made is that clocks do not measure ``time'' in the conventional sense, but rather indicate the variable ``time'' that they themselves generate. A distinction is drawn between ideal and actual clocks, with the understanding that any physical interactions do not affect the fundamental period of an ideal clock. Physical theories, with the exception of general relativity, assume the presence of ideal clocks. In Section \ref{inertialandnot}, we examine an operative criterion for differentiating between inertial and non-inertial reference frames alongside the traditional definition of an inertial reference frame. In this context, the accelerometer serves a critical role, noted for its remarkable feature of being unable to distinguish between acceleration and gravitational fields. Section \ref{motoaccsec} addresses the infinities arising from the Newtonian treatment of uniformly accelerated motion, highlighting the necessity for a velocity limit. The main section \ref{kineffects} and its subsections explore the kinematic effects of special relativity, including time dilation, length contraction, the luminous Doppler effect, Lorentz transformations, and the relativity of simultaneity. This theoretical analysis is complemented by a discussion of experimental validations of time dilation and the journey effect, which addresses the well-known clocks paradox. Section \ref{dynamics} focuses on a topic often overlooked in textbooks and educational practices: the emission and absorption of a photon by an atom or nucleus. In the context of Newtonian mechanics, this discussion leads to the concept of rest energy associated with mass, along with its implications. Section \ref{discussion} presents our proposal in conjunction with other methodologies, particularly those rooted in the space-time framework. This section also includes a presentation and analysis of preliminary tests conducted in two classroom settings. Finally, three appendices -- covering length contraction through the exchange of light flashes, Einstein's second thought experiment on simultaneity, and the gravitational red-shift as understood in the context of special relativity -- conclude the paper.
\section{About clocks}\label{clocksec}
It is often stated that ``time is measured by clocks'', much like ``electric current is measured by ammeters''. However, there is a fundamental difference between the two. An ammeter measures a property of something distinct: the electrical current in a circuit. The measurement involves an interaction between the ammeter and the circuit, which alters the circuit's current in the process. In contrast, clocks simply display their numbers or the positions of their hands; they do not measure anything external to them.
The numbers displayed on a clock, which are derived from a periodic phenomenon, result from a counting process that can be viewed as a form of measurement. However, this measurement relates specifically to the properties of the clock itself, rather than to something external to it.
\par
Therefore, {\em time must be understood as a mathematical variable that appears in equations, generated and indicated by a clock from an experimental standpoint}. For instance, imagine a plane graph paper that moves at a constant speed along the x-direction while a nib traces the measured value of a physical quantity along the y-direction. As the paper moves along the x-direction, we are not measuring anything; we are merely generating the values of the variable \( t \).
\par
On the other hand, a clock allows us to measure the duration of a phenomenon occurring in its vicinity (where the clock is). To do this, we must define the initial and final events of the phenomenon. At the moment the initial and final events occur, we read the times \( t_i \) and \( t_f \) displayed by the clock. By definition, the duration of the phenomenon is given by \( (t_f - t_i) \).

There are no specific guidelines for constructing a clock, except that the time values generated by the clock must accurately reflect the homogeneity of the mathematical variable \( t \). To fulfill this requirement, we compare different types of clocks and empirically assess which class performs best, guided by the theories that describe each class's operation.

We define an {\em ideal clock} as one whose fundamental period is not altered by any physical interaction.
The fact that the clocks used in experiments are actual clocks -- whose fundamental period is affected by many physical interactions -- constitutes a
problem that is left to experimenters.
In physical theories, clocks are considered {\em ideal clocks}.
The theory of general relativity constitutes an exception: this
theory requires taking into account the dependence of the fundamental period of
a clock on gravity, even at a theoretical level.
\par
\textsf{The discussion surrounding the role of clocks contrasts with the understanding that students have developed, which is primarily that clocks merely measure time. If students do not effectively clarify and articulate this fundamental concept, it will greatly hinder their comprehension of special relativity.
The idea that ``clocks measure time'' is a deeply rooted concept found in textbooks, teaching practices, and general intuition. This perspective suggests that ``time'' is something external to the clocks themselves. Instead, the way clocks display their numbers or hands implies that ``time'' is a distinct physical quantity represented by these devices.
As shown above, a more compelling argument can be made by considering a chart recorder, which tracks the value of a measurable physical quantity as a function of the variable time represented by the progression of the chart.}
\par
\textsf{Teachers should discuss in depth that, while clocks do not measure -- in the conventional sense -- time, we can measure the duration of a phenomenon with clocks. They must emphasize the significance of understanding the principal classes of clocks, particularly the fundamental clock period, which is the same in every IRS. In this context, it is vital to critically examine and challenge commonly heard statements, such as ``a moving clock runs slower''. This foundation is essential for a proper understanding of time dilation.}
\par
Every clock operates using a frequency standard, a counter, and a display. For example, in a pendulum clock, the frequency standard is the inverse of the pendulum's oscillation period. The counter is a mechanical device that converts the count of oscillations into the movement of the clock's hands. In a cesium clock, the frequency standard is determined by the frequency of the radiation absorbed by the two hyperfine sub-levels of the fundamental energy level of $^{133}$\,Cesium atoms. The counter in this case is electronic, and the display is digital.
In quartz clocks, the frequency standard is based on a vibrating piece of quartz. Similar to cesium clocks, the counter is electronic, and the display can be either analog or digital. The frequency standard used provides a cesium clock with better accuracy.
\par
Special attention should be given to pendulum clocks, which students typically encounter when studying Galileo's experiments. These clocks only function in a gravitational field, and their fundamental period is influenced by the gravitational field \(g\) as expressed in the formula \(T = 2\pi \sqrt{l/g}\), where \(l\) is the length of the pendulum. Consequently, a pendulum clock cannot operate in a laboratory that is in free fall (see section \ref{accesection}).
\section{Inertial and accelerated frames of reference}\label{inertialandnot}
A physical reference system is defined as a rigid body with an associated clock (or a system of synchronized clocks) and a set of coordinates. In literature and teaching practice, an inertial reference frame (IRS) is typically defined as follows: ``A reference system is said to be inertial if a body on which no net force is exerted is either at rest or moving in a straight line at a constant speed''.

This definition poses operational challenges because it is impossible to devise an experimental procedure to determine whether a force is acting on a body, other than through the observation that the body is at rest or in uniform rectilinear motion. Consider the following statement: if a body is at rest or its motion is uniform and rectilinear, then no force is acting on it. This statement is logically equivalent to the previous definition; however, it cannot be used to define an IRS \footnote{For an in-depth discussion of this issue, see, for instance, \cite{IRS} and references therein.}.

Despite its questionable epistemological status, this standard definition can still be useful for critically discussing the first two postulates of Newtonian mechanics.
\par
In the literature, we also find definitions like the following: ``Reference frames in which Newtonian mechanics holds are called inertial reference frames or inertial frames. Reference frames in which Newtonian mechanics does not hold are called non-inertial reference frames or non-inertial frames'' \cite[p. 94]{HR}. However, this definition is open to criticism. Rigorously speaking, Newton's dynamical law, given by $\vec F = m\vec a$, is not universally valid, as established by special relativity. It is important to specify that Galilean transformations are assumed to make this definition accurate. Under these transformations, Newton's dynamical law remains invariant.

Moreover, it is worth noting that Newtonian dynamics can also be applied in accelerated reference frames. Therefore, the definition mentioned above only holds when considering forces arising from interactions with other objects.

An operational definition of an inertial reference frame is: ``A reference system is inertial if its acceleration, {\em as measured within the same system}, is zero.'' This definition implies that if an accelerometer is used in a sufficiently small free-falling laboratory, it will indicate that the laboratory is in an inertial reference frame (see the next section). Conversely, a reference system {\em in which} the acceleration is not zero is classified as an accelerated system or as one situated within a gravitational field (see the next section).
\par
If there is one inertial system, there are countless others that move with uniform rectilinear motion relative to it. A laboratory located on Earth can be considered an inertial system with acceptable approximation, provided we ignore the Earth's gravitational and magnetic fields and assume the centripetal acceleration due to its rotation is sufficiently small.

Until recent decades, physics experiments have  been conducted in terrestrial laboratories. However, experiments have now also been carried out in laboratories situated in orbit around the Earth and in spacecraft in free fall.
\par
\textsf{The operative definition of inertial and accelerated reference systems should follow the traditional one while also exploring their connections and differences.
In this discussion, we utilized the concept of a field. Teachers should introduce this concept when explaining Newton's law of universal gravitation, emphasizing that the gravitational field, represented by \(\vec{g}\), has dimensions of acceleration.}
\subsection{How an accelerometer works}\label{accesection}
Accelerometers are commonly found in everyday devices, such as mobile phones and airbags. Teachers can use these familiar experiences to encourage students to investigate the behavior of accelerometers and uncover their remarkable properties.
\begin{figure}[htb]
\centering{
\includegraphics[width=5cm]{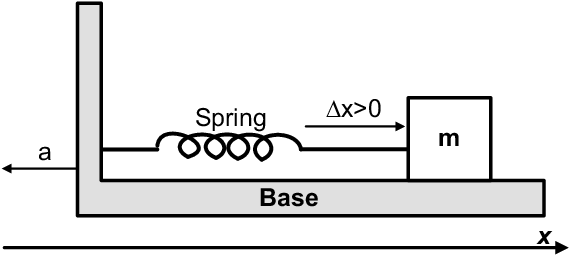}
}
\caption{\label{acce}  Working principle of an accelerometer. See the text.}
\end{figure}
\par\noindent
A mass $m$ is connected to a rigid base by a spring (Fig. \ref{acce}).
Ideally, the mass can slide on the base without friction. Suppose the base
is subjected to a constant acceleration $a$ to the left: the
spring is stretched, and its  maximum extension $\Delta x$ is
related to the acceleration of the base by the equation:
\begin{equation}\label{accelerometr}
\overrightarrow{\Delta x}=-\frac{m}{k} \vec a,
\end{equation}
where $k$ is the spring's constant. The elongation of the spring
occurs along the opposite direction of the base's acceleration.
\par
When the accelerometer is rotated 90 degrees to the right, it assumes a vertical position. In this position, the spring elongates downward due to the influence of the gravitational field \(\vec{g}\) acting on the mass \(m\), transforming the accelerometer into a gravimeter. Here, the accelerometer indicates an acceleration equal to \(-\vec{g}\), which is directed upwards. It is important to note that {\em an accelerometer does not differentiate between an acceleration field and a gravitational field}.

This observation implies that the effect of a gravitational field \(\vec{g}\) on a mass \(m\) is equivalent to the effect of an acceleration field \(\vec{a} = -\vec{g}\) on the same mass. Consequently, we can conclude that \(m_g = m_i\), where \(m_g\) represents the "gravitational mass" that appears in Newton's law of gravitation, and \(m_i\) signifies the "inertial mass" found in the dynamical equation \(\vec{F} = m_i \vec{a}\). In metric gravitation theories, the equivalence of inertial and gravitational mass is assumed as the "weak equivalence principle"\footnote{In special relativity, the
mass $m$ is no longer a measure of a body's inertia. Indeed, the
concept of inertial mass rests on using the equation $\vec F=m\vec a$, which is no longer valid in special relativity.}.
\par
From the above considerations, it follows that -- {\em in an accelerated reference frame} -- the measured acceleration $\vec a$ is equivalent to a pseudo-gravitational field $\vec g=-\vec a$. Hence,  a force $m\vec g=m(-\vec a)$ will be exerted on a mass $m$.
\par
\textsf{Students can be reminded of the physical sensations experienced by a passenger on a train.
If the train decelerates with a constant acceleration of \(-a\), the passenger feels a force of \(ma\) in the direction of the train's motion. Conversely, if the train accelerates with a constant acceleration, the passenger feels a force of \(ma\) in the opposite direction of the train's motion, which gives the sensation of walking uphill. }
\par
 \textsf{In an ideal experiment, the passenger is replaced by a cube that can glide without friction across the train floor.
In the initial setup, the train is moving with a velocity \( V \) along the railway while the cube is at rest on the train. Both the train and the cube are equipped with accelerometers. If the train decelerates with a constant acceleration of \(-a\), as measured by the train's accelerometer, the cube's accelerometer will register an acceleration of \(+a\) in the original direction of the train's movement. In this scenario, the train perceives the cube to be in uniform, accelerated motion.
Conversely, if the train accelerates, the cube will exhibit uniformly accelerated motion in the opposite direction of the train's movement. This phenomenon can be qualitatively observed on the train by using a suitcase equipped with wheels, which many students may have witnessed in real life.}
\par
\textsf{In the station's reference system, no forces act on the cube. Therefore the cube keeps moving with its velocity $V$, i.e., the train's velocity before its deceleration.
}
\par
If a laboratory, with an accelerometer fixed in the vertical position on a wall, is in free fall within a gravitational field, the spring remains unextended. This is because the gravitational force \( mg \) acting on the accelerometer's mass \( m \) is balanced by the pseudo-gravitational force \( -mg \) arising from the free-falling acceleration \( g \). Alternatively, both the accelerometer's mass and the wall to which the spring is attached fall together, thereby maintaining their relative distance (as per Galileo's reasoning; see below). Given that we define an inertial reference system as one in which the measured acceleration is zero, it follows that a free-falling laboratory represents an inertial reference frame. Consequently, a body that is released will either remain at rest or, if it possesses an initial linear momentum, continue to move uniformly in a straight line. However, this holds true only under the condition that the gravitational field is uniform, which it is not. Therefore, the earlier statement is approximately valid, provided that the dimensions of the laboratory are sufficiently small.
 \par
 \textsf{Consider a laboratory in free fall, orbiting in a circular path within the Earth's equatorial plane.
 Let $R$ be the distance of the laboratory's center of mass to the center of the Earth, and $R+a$ the distance of a point $P$ in the laboratory to the center of the Earth. Let also $h$ be the altitude of the orbit, such that $R=R_{Earth}+h$.
 The ratio of the gravitational field strength at the laboratory's center of mass to that at the point \( P \) is approximately given by \( 1 + {2a}/{R} \).
If we take \( h  = 400 \, \rm{km} \)  ($ 400\, \rm{km}$ is a typical altitude of the International Space Station) and \( a = 10 \, \rm{m} \), we find that \( {2a}/{R} \approx 3 \times 10^{-6} \) . This indicates that the center of mass perceives a mass located at \( P \) as experiencing an outward, radial acceleration, given by:
\begin{equation}\label{acc-iis}
    a_r\approx \frac{GM_{Earth}}{R^2}\frac{2a}{R},
\end{equation}
that yields $a_r\approx 3\times 10^{-5}\, {\rm ms^{-2}}$. It follows that, in one second, the mass moves over a distance of $
\approx 15 {\rm \mu m}$.
}
\par
Finally, let us consider a spaceship orbiting the Earth and a space module attempting to rendezvous with it. Successfully achieving this rendezvous requires a thorough understanding of orbital physics, particularly the tidal forces resulting from the gradient of the gravitational field, as discussed previously, and the effects of Coriolis forces \cite{RV}.

Coriolis forces are relevant in this context because orbiting space stations rotate about their axes (which are perpendicular to the plane of the orbit) at the same angular velocity at which they rotate around the Earth, allowing them to maintain a constant face towards the planet. For a more simplified analysis, refer to \cite{besson}.
\par
\textsf{Discussing free fall, teachers should read and discuss with students the following excerpt from Galilei's {\em Discorsi e dimostrazioni
matematiche intorno a due nuove scienze} \cite{galilei, galilei2}}:
\begin{quote}\small
\textsf{A large stone placed in a balance not only acquires additional
weight by having another stone placed upon it, but even by the
addition of a handful of hemp its weight is augmented six to ten
ounces according to the quantity of hemp. But if you tie the hemp
to the stone and allow them to fall freely from some height, do you
believe that the hemp will press down upon the stone and thus
accelerate its motion or do you think the motion will be retarded
by a partial upward pressure? One always feels the pressure upon
his shoulders when he prevents the motion of a load resting upon
him; but if one descends just as rapidly as the load would fall how
can it gravitate or press upon him? Do you not see that this would
be the same as trying to strike a man with a lance when he is
running away from you with a speed which is equal to, or even
greater, than that with which you are following him? You must
therefore conclude that, during free and natural fall, the small
stone does not press upon the larger and consequently does not
increase its weight as it does when at rest}.
\end{quote}
\textsf{Galilei's observations indicate that a free-falling body appears to ``lose its weight''. This conclusion aligns with the earlier result without the need for mathematical formulas. The qualitative aspects of a free-falling body can be demonstrated in the classroom using a simple device, as shown in \cite{bbggem}. This hands-on experiment will help students grasp the formal concepts discussed previously.}
\subsection{A puzzling infinite energy}\label{motoaccsec}
Suppose that a constant force $F_x$ operates on a mass $m$ at rest at the time $t=0$. In Newtonian mechanics, the acceleration of the mass will be:
\begin{equation}\label{accN}
a_x=\frac{F_x}{m}.
\end{equation}
This kind of motion is called uniformly accelerated motion.
The particle's velocity is given by:
\begin{equation}\label{velN}
v_x=a_xt.
\end{equation}
Hence, the kinetic energy of the particle is:
\begin{equation}\label{kinN}
E_k=\frac{1}{2}mv^2_x=\frac{1}{2}ma^2_xt^2.
\end{equation}
Eq. (\ref{velN}) says that  $v_x\rightarrow\infty$ when $t\rightarrow\infty$. Analogously, Eq. (\ref{kinN}) says that  $E_k\rightarrow\infty$ when $t\rightarrow\infty$.
The peculiarity of Eq. (\ref{kinN}) lies in the fact that applying an infinitesimally small force to an arbitrarily large mass can result in infinite kinetic energy. This strange outcome stems from Eq. (\ref{velN}), which permits infinite velocities. These observations strongly indicate that there should be a limit on the velocity of all material bodies. As a result, the principles of Newtonian dynamics may need to be revised.
\begin{figure}[h]
\centering{
\includegraphics[width=9cm]{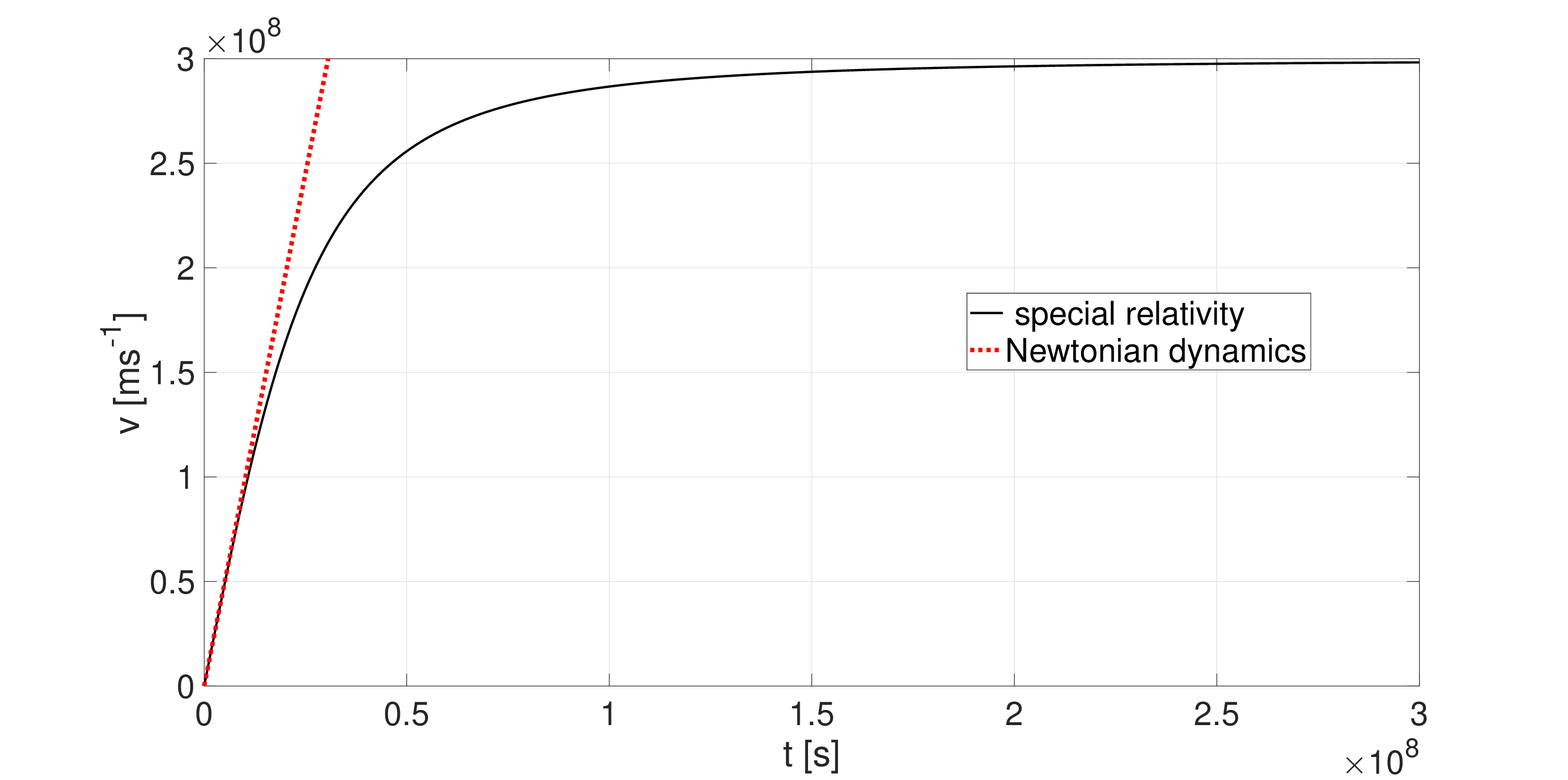}\\
}
\caption{Velocity of a particle under a constant force according to Newtonian and relativistic dynamics. In the relativistic case, the acceleration $a=dv/dt$ is not constant, i.e., the motion is not uniformly accelerated: see also Eq. (\ref{acc}) and Fig \ref{acc}. We have assumed $A=9.8\, \rm{ms^{-2}}$. See the text.}\label{mua}
\end{figure}
\par\noindent
In relativistic dynamics, Eq. (\ref{accN}) is replaced by the equation (we shall omit the suffix $x$):
\begin{equation}\label{eqmoto}
\frac{d( m \gamma v)}{dt}=F;\qquad \textrm{F\, constant},
\end{equation}
where $\gamma =1/\sqrt{1-v^2/c^2}$.
\begin{figure}[h]
\centering{
\includegraphics[width=8.5cm]{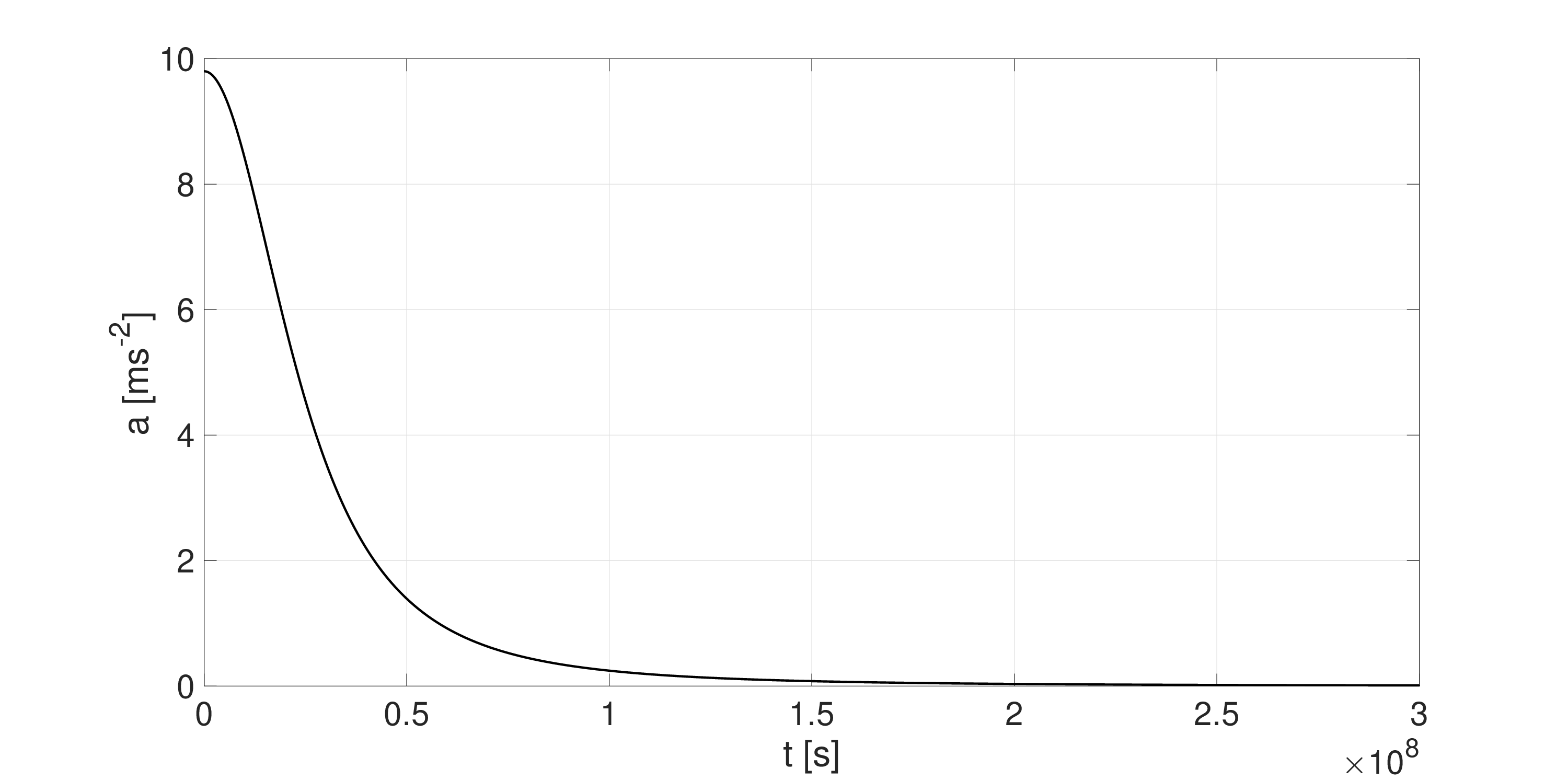}\\
}
\caption{Acceleration of a particle under a constant force. We have assumed $A=9.8\, \rm{ms^{-2}}$.}.\label{acc}
\end{figure}
\par\noindent
This equation can also be written as:
\begin{equation}\label{cosi}
\frac{d(\gamma v)}{dt}=\frac{F}{m} = A,
\end{equation}
where $A$ has the dimensions of an acceleration, and its value coincides with the Newtonian acceleration.
By integration, we get:
\begin{equation}\label{inte}
\gamma v=At+\gamma(0)v(0).
\end{equation}
Taking into account the initial conditions:
$v(0)=0;\, \gamma(0)=1$, we can write:
\begin{equation}\label{ci}
v=At\sqrt{1-\frac{v^2}{c^2}}.
\end{equation}
Squaring both members of this equation, we obtain:
\begin{equation}\label{vat}
\frac{v^2}{c^2}=\frac{A^2t^2}{c^2+A^2t^2}.
\end{equation}
Putting this value into Eq. (\ref{ci}), we finally get:
\begin{equation}\label{defini}
v=\frac{At}{\sqrt{1+A^2t^2/c^2}}.
\end{equation}
Hence, if $t\rightarrow\infty$, then $v\rightarrow c$. See Fig. \ref{mua}.
Deriving this equation with respect to $t$, we obtain the expression of the acceleration of the particle:
\begin{equation}\label{accelab}
a= \frac{dv}{dt}=\frac{Ac^3}{(c^2+A^2t^2)^{3/2}}.
\end{equation}
The graphic of this equation is drawn in Fig. \ref{acc}: the particle's acceleration goes to zero for sufficiently high values of $t$.
\par
Let us now consider the reference system instantaneously co-moving with the accelerated particle.
Using the transformation equation for the $x$ component of the force:
\begin{equation}\label{fx}
F_x  = {{F'_x+V/c^2(\vec F'\cdot \vec v\,')}\over{1+Vv'_x/c^2}},
\end{equation}
we obtain, in our case that $F'_x=F_x= \textrm{costant}$
because in Eq.
(\ref{fx}), $\vec v\,'=0$ e $v'_x=0$, since the particle is at rest in the reference frame, instantly co-moving with the particle.
Then the equation of relativistic dynamics reads:
\begin{equation}\label{din'}
F'_x=F_x=\frac{d(m\gamma' v'_x)}{dt'}.
\end{equation}
Therefore, since  $\gamma'=1$:
\begin{equation}\label{fine!}
a'_x= \frac{F_x}{m}= \textrm{costant}.
\end{equation}
We have thus concluded that the proper acceleration, namely the acceleration measured in the reference frame co-moving with the particle, is constant. Hence, we can define the reference system co-moving with a particle on which a constant force is applied as uniformly accelerated.
\par
\textsf{Only a portion of the treatment outlined above can be effectively incorporated into teaching elementary physics courses or high school classes. The extent and depth of this incorporation will depend on the specific teaching context. Teachers should emphasize the concept of ``puzzling infinite energy'' and the resulting necessity for a limited speed. Additionally, when introducing Newton's law of dynamics, expressed as \(\vec F = m\vec a\), educators may consider presenting it in an alternative form:
\begin{equation}\label{dp/dt}
\vec F =\frac{d\vec p}{dt},
\end{equation}
with $\vec p= m\vec v$. After that, teachers should state that, in relativistic dynamics, the fundamental law is:
\begin{equation}\label{dp/dtrel}
\vec F= \frac{d}{dt} (m\gamma \vec v); \qquad \gamma=\frac{1}{\sqrt{1-v^2/c^2}},
\end{equation}
and show Figs. \ref{mua} and \ref{acc}. The development of the intervening equations depends on the teaching context. Teachers should refer to the experiment conducted by Bertozzi, who measured the kinetic energy of electrons in the range of \(0.5 - 15 \, \rm{MeV}\) by analyzing the time of flight of electrons over a specific distance \cite{bertozzi}. The experimental results support the existence of a maximum velocity, denoted as \(c\). Beginning with Bertozzi's experiment provides an alternative approach to familiarizing students with the concept of a speed limit.
}
\section{Kinematic effects in special relativity}\label{kineffects}
The principle of relativity states that every phenomenon occurs in the same manner in every inertial reference system (IRS). In other words, the equations that describe a phenomenon maintain the same mathematical form in all IRS. This principle applies to Galilean and Newtonian physics, as well as to special relativity.

An immediate consequence of the principle of relativity is that the fundamental clock period remains constant across all IRS.
\par
The postulates of special relativity are:
\begin{enumerate}
\item Homogeneity of the time (of the time variable $t$)
\item Homogeneity and isotropy of space
\item Principle of relativity
\item The light speed in a vacuum is the same in every IRS\label{limitspeed}
\end{enumerate}
As recognized by Einstein \cite{does}, the fourth postulate is equivalent to: ``Maxwell's equations in a vacuum are valid''.
The first three postulates remain unchanged from those in Galilean and Newtonian relativity. Therefore, the key conceptual shift from previous physics lies in the fourth postulate: the constancy of the speed of light, or more specifically, the acknowledgment that the speed of light in a vacuum represents a universal speed limit.
\par
\textsf{Regarding the fourth postulate, teachers should remind students that, according to the principle of relativity, Maxwell's equations maintain the same mathematical form in every inertial reference system (IRS). Since these equations describe the propagation of light in a vacuum at a specific speed, that speed will remain constant in all IRS. Additionally, measurements of the speed of light conducted in any IRS will yield the same value.}
\par
As Einstein did, special relativity can be developed without introducing the concept of space-time \cite{ein05r}. It is helpful to derive all the predictions of relativistic kinematics -- such as time dilation, length contraction, and Lorentz transformations -- using thought experiments that involve the exchange of light flashes of null duration between two ideal clocks in relative inertial motion. In principle, these experiments can be conducted. Herman Bondi first suggested this approach to understanding the effects of relativistic kinematics in 1965 \cite{bondi}. For further reference, see \cite{ugarov}. The calculations rely on the times of flight for light pulses and the distances traveled by either a light pulse or a material body. These calculations require only simple algebraic manipulations, making them accessible to students in introductory physics courses or in upper high school years.
\subsection{The `$k$' method}\label{k-method}
\begin{figure}[htb]
\centerline{
\includegraphics[width=7.5cm]{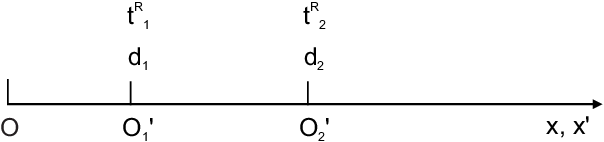}
} \caption{\label{lampi09} See the text.}
\end{figure}
\par\noindent
We consider two inertial frames, denoted as $K$ and $K'$. These frames are in relative motion, moving away from each other along the $x \equiv x'$ axis with a positive relative velocity $ V$.
Let $O$ and $O'$ be the origins of $K$ and $K'$.
$O$ and $O'$ are endowed with an ideal clock: the two clocks do not need to be synchronized.
$O$ emits two light flashes at the times $t_1$ and $t_1+T$. $O'$ will receive the two flashes at the times $t'_1$ and $t'_1+T'$, and instantaneously, it will reflect the two flashes towards $O$.
\par
For the homogeneity of the variable
time, the value of $T'$ cannot depend on the time $t_1$; for the homogeneity
of space, $T'$ cannot depend on the distance between $O$
and $O'$ when $O$ emits the two flashes. \label{k-def-page}Hence, $T'=  kT$, where
$k > 0$ is a parameter to be determined. {\em $k$ must be positive; otherwise $O'$	
would receive the second flash before the first}.
\par
If $t^*_1$ and $t^*_2$ are the times
at which $O$ receives the two flashes reflected by $O'$, we have, for
the space's isotropy and the relativity principle, that:
\begin{equation}\label{kappa}
t^*_2-t^*_1=kT'=k^2T.
\end{equation}
If $d_1$ and $d_2$ are the distances at which, according to $O$, $O'$ receives and
reflects the first and second flash (Fig. \ref{lampi09}), then:
\begin{equation}\label{distanze}
2d_1= c(t^*_1-t_1); \qquad 2d_2= c(t^*_2-t_2).
\end{equation}
Therefore:
\begin{equation}\label{diffedist}
(d_2-d_1)= \frac{c}{2}[(t^*_2-t^*_1)
-(t_2-t_1)]=\frac{c}{2}(k^2T-T)=\frac{cT}{2}(k^2-1).
\end{equation}
This equation yields the difference $ (d_2-d_1)$ in the parameter $k$.
\par
The same difference can be calculated in another way.
If the times at which $O'$ reflects the two flashes are $t^R_1$ e $t^R_2$ (Fig. \ref{lampi09}), we have:
\begin{equation}\label{riri}
t^R_1= t_1+ \frac{d_1}{c};\qquad t^R_2=(t_1+ T)+\frac{d_2}{c}.
\end{equation}
From the above two equations, we get:
\begin{equation}\label{tR2-tR1}
t^R_2-t^R_1=T+\frac{d_2-d_1}{c}.
\end{equation}
On the other hand, $t^R_2-t^R_1$ can be written as:
\begin{equation}\label{diffdist1}
t^R_2-t^R_1=\frac{(d_2-d_1)}{V}.
\end{equation}
Equating the second members of these two equations, we obtain:
\begin{equation}\label{diffedist2}
(d_2-d_1)=\frac{VT}{1-B};\qquad {\rm with}\:B=\frac{V}{c}.
\end{equation}
Equating the second members of Eqs. (\ref{diffedist}) and (\ref{diffedist2}), we get:
\begin{equation}\label{k2}
k^2=\frac{1+B}{1-B}.
\end{equation}
Namely, since $k>0$:
\begin{equation}\label{kappa09}
k=\sqrt{\frac{1+B}{1-B}}=\sqrt{\frac{(1+B)(1+B)}{1-B^2}}=\Gamma (1+B),
\end{equation}
where:
\begin{equation}\label{Gamma}
\Gamma=\sqrt{\frac{1}{1-B^2}},
\end{equation}
is, as we shall see, the time dilation factor.
From Eq. (\ref{kappa09}) we see that $k\ge 1$, where the equality holds if $B=0$, i.e., if $V=0$ \footnote{Notice the different notations for the ratio $\beta=v/c$ and $B=V/c$; similarly for $\gamma$ and $\Gamma$. $\beta$ and $\gamma$ contain the velocity $v$ of a particle while $B$ and $\Gamma$ contain the relative velocity $V$ between two IRS.}.
\par
Furthermore, it follows that:
\begin{eqnarray} \label{kk}
{{k ^2+1} \over {2 k }} & = & \Gamma  \\
&&\nonumber\\
{{k ^2 -1} \over {2 k }} & = & \Gamma  B
 \end{eqnarray}
If $O$ and $O'$ are approaching each other along the common $x\equiv x'$ axis at constant velocity $-V$, we should carry out the calculations from the start: everything goes as before by substituting $ V $ with $ -V $ in every formula. Hence, while the expression of $\Gamma$ remains unchanged, the factor $k$ assumes the value:
\begin{equation}\label{k-v}
k(-V)=\Gamma (1-B).
\end{equation}
This last equation will be helpful when discussing the light Doppler effect.
\par
\textsf{Teachers should note that the `$\kappa$ factor' depends only on the relative velocity between the two clocks, while the $\Gamma$ factor can be quickly derived from the simple calculations mentioned. This method of exploring relativistic kinematic effects encourages students to engage in activities involving the sending and receiving of light signals. Such exercises will help students move away from the notion that, within a given inertial reference frame (IRF), the time associated with an event is the same as the time the observer receives the signal. Refer to point \ref{signal} on page \pageref{signal} and see \cite{simul} for further details.
}
\subsection{Time dilation}\label{TDsec}
Let us now consider the phenomenon whose initial and final events are
the reflection, by $O'$, of the first and second flash.
The duration of this phenomenon, according to $O$, is given by (using the Eqs.
(\ref{diffdist1}) and (\ref{diffedist2})):
\begin{equation}\label{duratalunga}
\Delta t= t^R_2-t^R_1=\frac{T}{1-B}.
\end{equation}
Instead, the duration of the same phenomenon according to $O'$ is:
\begin{equation}\label{duratapropria09}
\Delta t'=T'=kT.
\end{equation}
From this equation, we get:
\begin{equation}\label{Tek}
T=\frac{\Delta t'}{k}=\frac{1}{\Gamma}\frac{1}{1+B}\Delta t',
\end{equation}
where we have used Eq. (\ref{kappa09}).
Substituting this value of $T$ into Eq. (\ref{duratalunga}), we obtain at the end:
\begin{equation}\label{radar2}
\Delta t = \Gamma \Delta t'={{\Delta t'}\over{\sqrt{1-V^2/c^2}}}.
\end{equation}
The phenomenon develops at the same point in $O'$. The duration of a phenomenon that develops at the same point of an IRS is called {\em proper duration}. Instead, according to $O$, the initial and final events of the phenomenon occur at different points of its IRS. As seen from Eq. (\ref{radar2}), the proper duration of a phenomenon is the shortest.
\par
In the above calculations, we have considered the phenomenon consisting of the reflection by $O'$ of the first and the second light signal sent by $O$. The duration of this phenomenon is measured in the reference frame of \( O' \) and calculated in the reference frame of \( O \). It is important for teachers to emphasize that time dilation is a symmetrical effect. We have demonstrated that if \( O' \) informs \( O \) about the duration of a phenomenon occurring in \( O' \), the result is given by the equation \( \Delta t = \Gamma \Delta t' \) (see Eq. (\ref{radar2})).
\textsf{Now, let us switch the roles of \( O \) and \( O' \): when \( O \) informs \( O' \) about the duration of a phenomenon that occurs at \( O \), the relationship between the two durations changes to \( \Delta t' = \Gamma \Delta t \). This conclusion is consistent with the principle of relativity.}

\textsf{If students find this argument unconvincing, encourage them to redo the entire calculation while reversing the roles of \( O \) and \( O' \). They will discover that this lengthy recalculation can be avoided by simply applying the principle of relativity. For difficulties grasping this symmetrical property of time dilation, see \cite{timedilation1}, where the study's sample was pre-service teachers.}

\subsection{The Doppler effect}\label{DEsec}
Using the expression (\ref{kappa09}) found for the coefficient $k$ in the case
in which the two reference systems are moving away, we obtain:
\begin{equation}\label{doppler}
T'= T \sqrt {{{1+ B } \over {1- B }} },
\end{equation}
i.e.,:
\begin{equation}\label{dopplerr}
T'= T  {{{1+ {{V}/{c}} } \over {\sqrt{1-V^2/c^2}}} }.
\end{equation}
In the reference system of $O$, $T$ is the time interval between the emission of the first and second flash. Instead,
in the reference system of $O'$, $T'$ is the time interval between the reception of the first and the second flash.
If $O$ sends flashes of light at regular intervals $T$, then $O'$ will receive these flashes at regular intervals $T'$.
Hence, $T$ is the period of the signal constituted by the set of flashes sent by $O$, and $T'$ is the period of the signal constituted by the set of flashes received by $O'$. It follows that Eq. (\ref{dopplerr}) is the formula for the Doppler effect of light in a vacuum, a formula that can be written also in terms of frequencies:
\begin{equation}\label{dopplerf}
\nu\,' =\nu {{\sqrt {1 - V^2/c^2} \over {1+ V /c}}}.
\end{equation}
The frequency of the periodic signal received by $O'$ decreases if the two reference systems are flying apart. Notice that Eq. (\ref{dopplerr}) and (\ref{dopplerf}) are also valid for an electromagnetic wave: in this case, the signal is constituted by the set of the maximum values of the electric or the magnetic field of the wave.
\par
If $O$ and $O'$ are approaching instead of flying away from each other, we get, based on Eq. (\ref{k-v}):
\begin{equation}\label{doppleravv}
T'= T \sqrt {{{1- B } \over {1+ B }} } =  T  {{{1- {{V}/{c}} } \over {\sqrt{1-V^2/c^2}}} },
\end{equation}
and:
\begin{equation}\label{dopplerfa}
\nu\,' =\nu {{\sqrt {1 - V^2/c^2} \over {1- V /c}}}.
\end{equation}
We are comparing the Eq. (\ref{dopplerr}) with the Eq. (\ref{doppleravv}). The last equation can be obtained directly from the former by substituting $V$ with $-V$.
\par
\textsf{Teachers should note that the result mentioned above was obtained under a specific configuration: the two reference frames are either moving in the direction of signal propagation or in the opposite direction. A more general Doppler effect equation will be discussed in section \ref{dopp-pho-sec}. Additionally, students have previously studied the acoustic Doppler effect, which occurs only in a medium, typically air. This may lead students to believe that the acoustic and luminous Doppler effects are entirely different phenomena treated separately. However, both effects can actually be explained within a unified framework based on exchanging signals of ideally null duration. Refer to \cite{frankl, natura} and also \cite{doppler1}, where waves replace discrete signals.
The calculations are performed in the laboratory's reference frame, where the medium remains at rest. The formulas derived are the classical ones. To obtain the corresponding relativistic equations, we express them in the reference frames of both the emitter and absorber, incorporating time dilation effects. Ultimately, the ratio of the proper frequencies of the emitter and absorber results in the relativistic Doppler formula for both acoustic and luminous signals in a medium. For the luminous Doppler effect in a vacuum, we simply substitute the velocity of the signal with \(c\). All calculations are within the grasp of students in elementary physics courses or the upper years of high school.
}
\subsection{Length contraction}\label{LCsec}
If the speed of light in a vacuum is known, we can measure the distance between
two points $A$ and $B$ using a clock placed in $A$ and a mirror placed in
$B$; the mirror's function is to reflect the flash of light coming from $A$ instantly. Then, the distance $l_0$ between
$A$ and $B$ is given by:
\begin{equation}\label{lunpropria}
l_0=\frac{1}{2}   c\Delta t,
\end{equation}
where $\Delta t$ is the time interval, measured by the clock in $A$,
between the start of the flash and its return. Of course, we used
implicitly the postulate of the isotropy of space for which the velocity
of light is the same in the outward and return journey.
The same procedure can be applied, with some modifications, to
measure the length of a moving stick. The calculation is somewhat lengthy. See, Appendix \ref{LC}.
\par
\begin{figure}[h]
\centering{
\includegraphics[width=5cm]{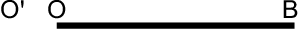}\\
}
\caption{The clock $O'$ and the stick $OB$ approach in relative, head-on inertial motion with relative velocity $V$.   Both reference frames consider the relative velocity $V$ as positive in their calculations.}\label{piattaforma}
\end{figure}

\textsf{If teachers consider that the calculation of Appendix \ref{LC} might be somewhat dull for their students, they can resort to the following shortcut. Referring to Fig. \ref{piattaforma}, we consider the phenomenon whose initial event is  ``O' meets $O$'', and whose final event is ``O' meets $B$''.
}
\par\noindent
\textsf{This phenomenon occurs at the same point $O'$. Therefore, its duration is the proper duration of the phenomenon. We have:
\begin{equation}\label{contra-O'}
    \Delta t'=\frac{l'}{V},
\end{equation}
where $l'$ is the length of the stick $OB$ evaluated by $O'$. The duration of the same phenomenon for the stick $OB$ is given by:
\begin{equation}\label{contra-OB}
    \Delta t=\frac{l_0}{V},
\end{equation}
where $l_0$ is the proper length of the stick.
Dividing member by member Eq. (\ref{contra-OB}) and Eq. (\ref{contra-O'}), we get, since $\Delta t=\Gamma \Delta t'$:
\begin{equation}\label{contra-contra}
    \frac{\Delta t}{\Delta t'}=\frac{l_0}{l'}.
\end{equation}
It follows immediately that:
\begin{equation}\label{Contra-contra-finale}
    l'=\frac{l_0}{\Gamma}=l_0\sqrt{1-\frac{V^2}{c^2}},
\end{equation}
i.e., the formula of length contraction. This calculation has the merit of showing at a glance that the length contraction is a corollary of the time dilation. However, it misses the operational value of the proof of Appendix \ref{LC}.
}
\subsection{The Lorentz transformations}\label{LT}
\begin{figure}[h]
\centering{
\includegraphics[width=7cm]{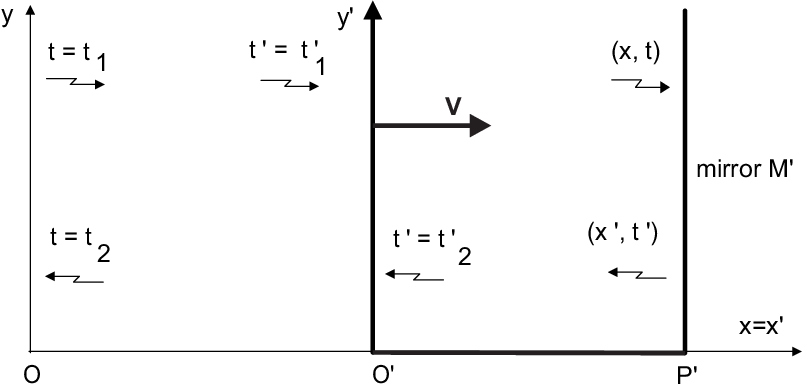}
} \caption{\label{lorentz} Thought experiment for deriving Lorentz transformations. See the text.}
\end{figure}

Consider the Fig. \ref{lorentz}. $O'$ is moving with velocity $V$ relative to $O$ along the positive direction of the common $x\equiv x'$ axis. We suppose that at the time $t=t'=0$ the positions of $O$ and $O'$ coincide at $x=x'=0$, and that at $t=0$ $O$ sends a light flash towards $O'$: this flash is received and reflected by  $O'$ at the time $t'=0$, and this reflected flash is received by $O$ at $t=0$. The mirror $M'$ is at rest in the reference frame $O'$. At the time $t=t_1$, $O$ sends a {\em second} light flash towards the mirror. The flash is received and retransmitted towards the mirror $M'$ by $O'$ at the time $t'_1$.The mirror receives and reflects the flash at the time $t$ and at the position $x$. For $O'$ this event occurs at the position $x'$ and at the time $t'$. The returning flash is received and retransmitted towards $O$ by $O'$ at the time $t'_2$ and reaches $O$ at the time $t_2$.
\par\noindent
We have:
\begin{eqnarray}\label{t12}
t_1& =& t- {{x} \over {c}} \label{t1}\\
t_2 & = & t+ {{x} \over {c}}. \label{t2}
\end{eqnarray}
Similarly, for $O'$:
\begin{eqnarray}\label{tt}
t'_1& = & t'- {{x'} \over {c}}  \nonumber\\
&&\\
t'_2 & = & t'+ {{x'} \over {c}}.\nonumber
\end{eqnarray}
Summing member by member the last two equations, we obtain:
\begin{equation}\label{somma}
t'= {{1} \over {2}} (t'_1+t'_2).
\end{equation}
On the basis of the definition of the coefficient $k$ at page \pageref{k-def-page}, we can write:
\begin{eqnarray}\label{kt}
t'_1-0 & = & k(t_1-0)  \\
&&\nonumber\\
t_2-0& = & k(t'_2-0).
\end{eqnarray}
Then, considering the equations (\ref{t1}),
(\ref{t2}) and (\ref{kk}), Eq. (\ref{somma}) assumes, after some algebraic manipulation, the form:
\begin{equation}\label{tem}
t' = \Gamma \left(t - {{B} \over {c}} x   \right)=\Gamma\left(t-\frac{V}{c^2}x \right).
\end{equation}
Instead, subtracting member by member the equations (\ref{tt}) and following the same steps, we obtain:
\begin{equation}\label{lun}
x' = \Gamma (x-Vt).
\end{equation}
These last two equations are the Lorentz transformations for the variables $x$ and $t$.
\begin{figure}[h]
 \centering{
\includegraphics[width=5cm]{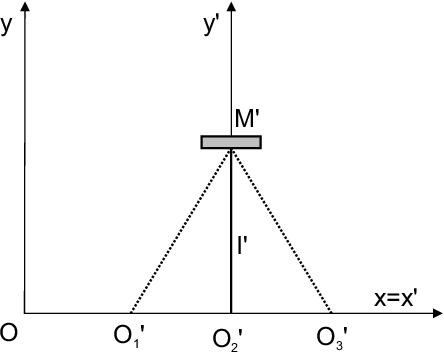}
 }
 \caption{\label{lt}  A light flash sent by $O'$ along the
$y'$ axis is reflected by the mirror
$M'$. The dashed lines represent the flash's path according to $O$. $O'_1,O'_2,O'_3$ are the successive positions of $O'$ in the reference frame of $O$. We have:
$\overline{O'_1O'_2}=\overline{O'_2O'_3}=
 V\Delta t/2$, where $\Delta t$ is the duration of the flash's round trip in the reference frame of $O$.}
 \end{figure}
\par\noindent
To get the transformation for the coordinate $y$ we proceed as follows (Fig. \ref{lt}).
 $O'$ sends a light flash along the positive direction of its $y'$ axis (that is parallel to the $y$ axis of $O$) towards the mirror $M'$ positioned at the distance $l'$. We have:
\begin{equation}\label{ltrasv}
 \Delta t'  = {{2l'}\over{c}},
\end{equation}
where $\Delta t'$ is the duration of the round trip of the flash.
$O$ describe the same phenomenon as illustrated in Fig. \ref{lt}. In the reference frame of $O$, the duration of the round trip of the flash is:
\begin{equation}
 \Delta t = {{2}\over{c}} \sqrt{l^2 + V^2{{\Delta t ^2}\over{4}}}.
\end{equation}
Squaring both members of this equation and rearranging the terms yields:
\begin{equation}
\Delta t^2\left(1-{{V^2}\over{c^2}}  \right)= \frac{\Delta t^2}{\Gamma^2}={{4}\over{c^2}} l^2.
\end{equation}
Finally, using the time dilation formula Eq. (\ref{radar2}) and Eq.
(\ref{ltrasv}), we get:
\begin{equation}
 l=l',
\end{equation}
namely $y=y'$. In the same way we can prove that $z=z'$.
\par
\textsf{The derivation of Lorentz transformations is lengthy. Usually, teachers write them and compare them with the Galilean ones. The derivation of $y'=y$ and of $z'=z$ could be carried on. Indeed, the thought experiment of Fig. \ref{lt} is often used in textbooks and teaching practices for presenting the ``light clock''.
}

\subsection{Relativity of simultaneity}\label{train}
The relativity of simultaneity can be explored through exercises involving Lorentz transformations. This method entails writing the time and space coordinates of two events in two inertial reference frames. While this approach is formal and does not require deep physical reasoning, it often falls short in educational contexts. Typically, textbooks and teaching practices address the relativity of simultaneity qualitatively or through simple calculations. However, there is still a need for a comprehensive quantitative treatment on the topic.
\par
The following topic will be examined using equations such as $x = a \pm V \Delta t$ and the formulas for time dilation and length contraction.
\begin{figure}[h]
\centering{
\includegraphics[width=7cm]{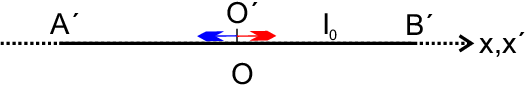}\\
}
\caption{\small{
The train $A'B'$ of proper length $l_0$ is in inertial motion with velocity $V$ along the positive direction of the $x\equiv x'$ axis. The train is coming from negative values of $x$. When $O'$ (center of the train) meets $O$ (center of the station's platform), their clocks are synchronized at $t=t'=0$, and $O'$ sends two flashes of null duration toward the train's head and tail.
The colors \textcolor[rgb]{1.00,0.00,0.00}{Red} e \textcolor[rgb]{0.00,0.00,1.00}{Blue} have only a graphical meaning.}}\label{rel-sim}
\end{figure}
\par\noindent
We shall discuss a well-known thought experiment (Fig. \ref{rel-sim}) illustrated in some textbooks (with variants conceptually irrelevant). Einstein and Infeld discussed the thought experiment of Fig. \ref{rel-sim}, where the train was replaced by a moving room \cite[pp. 187-188]{eininf}.
\par
According to $O'$, the flashes reach $A'$ and $B'$ simultaneously at the times $t'_{A'}=t'_{B}=l_0/2c$,
$l_0$ being the train's proper length. Indeed, $O'$ is mid-way between the train's head and tail, and for the isotropy of space, the light velocity is the same in both directions.
\par
Instead, $O$ describes the phenomenon as follows.
The motion of $A'$ (tail of the train), between $t=t'=0$ and  $t=t_{A'}$ is described by the equation:
\begin{equation}\label{motoA'}
x_{A'}= -\frac{l}{2}+Vt_{A'},
\end{equation}
where $l$ is the train length evaluated by $O$, and $t_{A'}$ is the time when the \textcolor[rgb]{0.00,0.00,1.00}{Blue} flash reaches $A'$.
$x_{A'}$ can also be written as:
\begin{equation}\label{lampoblu}
x_{A'}=-ct_{A'}.
\end{equation}
From these two equations, we get, by a simple calculation:
\begin{equation}\label{uguale}
t_{A'}=\frac{l}{2}\frac{1}{c+V}.
\end{equation}
Let us now consider the \textcolor[rgb]{1.00,0.00,0.00}{Red} flash.
Proceeding as in the case of the  \textcolor[rgb]{0.00,0.00,1.00}{Blue} flash we write:
\begin{equation}\label{motoB'}
x_{B'}= \frac{l}{2}+Vt_{B'},
\end{equation}
and:
\begin{equation}\label{anche}
x_{B'}=ct_{B'}.
\end{equation}
From these two equations, we obtain:
\begin{equation}\label{tB'}
t_{B'}=\frac{l}{2}\frac{1}{c-V}.
\end{equation}
Subtracting member by member Eq. (\ref{uguale}) from Eq. (\ref{tB'}), we get:
\begin{equation}\label{nosimul}
t_{B'}-t_{A'}=\frac{l}{2}\left(\frac{1}{c-V} -\frac{1}{c+V} \right)=\Gamma ^2B\frac{l}{c}.
\end{equation}
Since $l=l_0/\Gamma$ (length contraction), the final form of this equation is:
\begin{equation}\label{nosimul2}
t_{B'}-t_{A'}=\Gamma B\frac{l_0}{c},
\end{equation}
where $l_0$ is the train's proper length.
Thus, we have proved that the  events `arrival of the \textcolor[rgb]{1.00,0.00,0.00}{Red} flash at the head of the train' and `arrival of the \textcolor[rgb]{0.00,0.00,1.00}{Blue} flash at the tail of the train' are simultaneous for the train's center; instead, they are not for the center of the station's platform.
\par
Eq. (\ref{uguale}) can also be written as:
\begin{equation}\label{uguale2}
t_{A'}=\frac{l}{2}\frac{1}{c+V}=\frac{l}{2c}\frac{1}{1+B}=\frac{l}{2c}\frac{1-B}{1-B^2}=\frac{l}{2c}\Gamma ^2(1-B)=\frac{l_0}{2c}\Gamma (1-B),
\end{equation}
and Eq. (\ref{tB'}) as:
\begin{equation}\label{analogotB'}
t_{B'}= \frac{l_0}{2c}\Gamma (1+B).
\end{equation}
Naturally, the difference $(t_{B'}-t_{A'})$ remains unchanged.
\par
\textsf{The calculations above prompt students to write the equation of motion for a point moving with a constant velocity \( V \) along the \( x \) axis, as well as to determine the distance covered by a flash of light within a specified time interval. Students are then required to solve a simple system of two linear equations.}
\par
\textsf{
We have intentionally introduced length contraction only at the end of this discussion because it is essential for students to understand that the length of an object can be evaluated differently by two inertial reference systems (IRS). This scenario highlights the relativity of simultaneity and aims to give students a clear understanding of this kinematic effect in special relativity. In certain teaching contexts, instructors may wish to address the more complex case suggested by Einstein in his renowned book on special and general relativity \cite[pp. 30-33]{einrsrg}. We explore this case further in Appendix \ref{fulmini}.
}
\subsection{Using a $(x,ct)$ plane}\label{minksec}
\begin{figure}[h]
\centering{
\includegraphics[width=6cm]{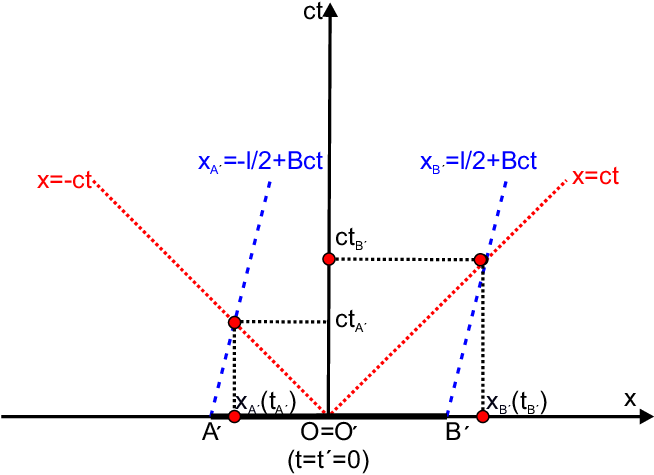}\\
}
\caption{Graphic representation in the plane $(x, ct)$ of the thought experiment of Fig. \ref{rel-sim} in the platform's reference frame. The line $A'B'$ is the train length, as seen from the platform. The dotted red lines represent the propagation of the light flashes from $O'$ towards $A'$ and $B'$. The dashed blue lines represent the motion of the train's ends $A'$ and $B'$ with velocity $V=Bc$. The intersections of the red and blue lines yield the positions and the times at which the train's ends receive the light flashes.
Looking at this figure, we see at a glance that: $t_{B'}>t_{A'}$; $ct_{B'}=x_{B'}(t_{B'}); ct_{A'}=-x_{A'}(t_{A'})$.}\label{mink}
\end{figure}
\par\noindent
The same thought experiment can be illustrated in a plane $(x, ct)$ (Fig. \ref{mink}). Since the equation of the motion of the light flash emitted in the positive direction of the axis $x \equiv x'$ (from $O'=O$ to $B'$) is $x=ct$, its representation in the plane $(x, ct)$ is the bisector of the right angle between the two axes.
Instead, the bisector of the second quadrant represents the equation $x=-ct$, which describes the motion of the light flash emitted in the negative direction of the axis $x \equiv x' $ (from $O'=O$ to $A'$).
The motion of the train's ends in the IRS of the platform is represented by two parallel lines whose equations are $x_{A'}=-l/2+Bct$ and  $x_{B'}=l/2+Bct$ where $l$ is the length of the train estimated by the platform and $B=V/c$. Therefore:
\begin{enumerate}
\item the  time $t_{B'}$ in which the flash towards $B'$ reaches $B'$ is given by the intersection of the two lines $x=ct$ and $x_{B'}=l/2+Bct $ \label{one}
\item the time $t_{A'}$ in which the flash towards $A'$ reaches $A'$ is given by the intersection of the two lines $x=-ct$ and $x_{A'}=-l/2+Bct$ \label{due}
\end{enumerate}
From point \ref{one} we deduce that $t_{B'}$ must satisfy the equation:
\begin{equation} \label{unoeq}
ct_{B'}=\frac{l}{2}+Bct_{B'}.
\end{equation}
Solving this equation for $t_{B'}$ gives:
\begin{equation} \label{unoeqris}
t_{B'}= \frac{l}{2c} \frac{1}{1-B}= \frac{l}{2c} \Gamma^2(1+B)= \frac{l_0}{2c} \Gamma (1+B),
\end{equation}
where $l_0$ is the train's proper length.
\par
Proceeding in the same way for $t_{A'}$, we get:
\begin{equation} \label{dueeqris}
t_{A'}= \frac{l_0}{2c} \Gamma (1-B).
\end{equation}
Comparing these two last equations it is found that $t_{B'}>t_{A'}$ and that:
\begin{equation} \label{diff1-2}
t_{B'}-t_{A'}= \frac{l_0}{c} \Gamma B.
\end{equation}
The position at which $B'$ receives the flash of light is given by:
\begin{equation} \label{posB'}
x_{B'}(t_{B'})=(t_{B'}-0)c=\frac{l_0}{2}\Gamma (1+B).
\end{equation}
Similarly, we obtain:
\begin{equation} \label{posA'}
x_{A'}(t_{A'})=-(t_{A'}-0)c=-\frac{l_0}{2}\Gamma (1-B).
\end{equation}
\textsf{The description of the thought experiment of Fig. \ref{rel-sim} in a plane $(x,ct)$ allow students to
visualize the main features of the thought experiment.
}
\subsection{The round-trip journey effect}\label{journeysec}
\textsf{The so-called clocks paradox (or twin paradox) is well known. In reality, there is no paradox at all, but only a round-trip journey effect.
This subject is commonly covered in elementary physics courses or high school, often sparked by students' curiosity about orbiting spacecraft and science fiction movies.
}
\par
In discussing this issue, we will demonstrate that time dilation and the journey effect are connected yet distinct phenomena.
To begin, let us restate the paradox. Two ideal clocks, \( O_1 \) and \( O_2 \), are initially at rest at the origin \( O \) of an inertial reference system (IRS), such as a laboratory's IRS. At a certain moment, \( O_2 \) accelerates in the positive direction of the \( x \)-axis until it reaches a velocity \( V \). After traveling at this velocity \( V \) for a certain interval of time, \( O_2 \) decelerates, stops at point \( B \), and then returns to \( O_1 \) following the same stages and methods used during the outward journey, ultimately rejoining \( O_1 \).
\par
Since these clocks are ideal, their fundamental period does not depend on any external fields, particularly acceleration fields. Therefore, we can consider acceleration and deceleration to be instantaneous. Furthermore, at the initial moment, both clocks were synchronized.
\par
When \( O_2 \) returns, \( O_1 \) predicts that it will show a lower reading than its own. However, since the motion is relative, \( O_2 \) also predicts that the reading on \( O_1 \) will be lower than its own. This creates the paradox.
\par
The paradox arises from the following flawed argument: since motion is relative (premise), then \( O_2 \) predicts that when they reunite, \( O_1 \) will show a lower elapsed time (conclusion). This conclusion does not logically follow from the premise because the two clocks are not equivalent.

In fact, while \( O_1 \) observes two events -- ``\( O_1 \) meets \( O_2 \)'' and ``\( O_1 \) meets \( O_2 \)'' for the second time -- \( O_2 \) observes three events: ``\( O_2 \) meets \( O_1 \)'' for the first time, ``\( O_2 \) meets \( B \)'', and finally, ``\( O_2 \) meets \( O_1 \)'' a second time.

Generally, the paradox is resolved as follows: the two clocks are not equivalent; in this case, only \( O_2 \) undergoes acceleration. Therefore, \( O_2 \) is the clock that will show the lower elapsed time upon reunification. It is also noted that while acceleration allows us to distinguish between the two clocks, the effect of time dilation is only due to the relative speed \( V \) between them during their periods of inertial motion.\par
Let us now consider three ideal clocks in the laboratory's IRS.
The clocks $O$, $O_1$ and
$O_2$ are at rest at the origin of the  $x$ coordinate. At a certain time, assumed equal to zero for all three clocks,  the clocks $O_1$ and $O_2$ begin a journey identical to that of the usual paradox, $O_1$ along the positive direction of $x$, $O_2$ along the negative direction.
As in the case of the standard paradox, we assume that the accelerations are instantaneous.
The inversion of the motion of $O_1$ occurs when $O_1$ reaches the point $B_1$ whose distance from $O$ is $l_0$.
The inversion of the motion of $O_2$ occurs when $O_2$ reaches the point $B_2$ whose distance from $O$ is $l_0$.
This configuration was discussed in \cite{terzetto} and was called
`the triplet paradox'.
\par
According to $O$, the clock at rest in the laboratory, the duration of the round-trip journeys of the two flying clocks is the same and equal to $2l_0 /V$: when the two clocks rejoin with $O$, they are still synchronized.
Also, $O_1$ and $O_2$ agree on this conclusion. Indeed, for $O_1$, the proper duration of its round-trip journey equals $(2l_0/V)\sqrt{1-V^2/c^2}$. Also, for $O_2$, the proper duration of its round-trip journey equals $(2l_0/V)\sqrt{1-V^2/c^2}$. Hence, when the two flying clocks rejoin, they are still synchronized.
However, the proper duration of the two round-trip journeys is lower by a factor $\sqrt{1-V^2/c^2}$ of the duration measured by the clock at rest in the laboratory. There is no journey effect if we compare the flying clocks $O_1$ and $O_2$. However, the journey effect is still there if we compare the flying clocks with the clock $O$ at rest in the laboratory.
\par
During their inertial flights, the relative velocity of the two flying clocks, $O_1$ and $O_2$, is $w=
2V/(1+V^2/c^2)$. Suppose that $O_2$ wants to inform $O_1$, by exchange of light signals,  about the duration $\Delta t_2$ of a phenomenon that develops at $O_2$.
Then, we shall have  $\Delta t_1=\Gamma \Delta t_2$, with $\Gamma =1/\sqrt{1- w^2/c^2}$. And vice versa. Hence, $O_1$ and $O_2$ register the time dilation effects but not the round-trip journey effect. So, we have proved that time dilation and the journey effect are connected but different phenomena.
\par
An asymmetrical variant of the triplet paradox -- from now on denoted as the triplet journey effect -- not considered in \cite{terzetto}, is particularly interesting.
Now the distances $l_1$ and $l_2$ from $O$ covered by the two flying clocks are different; different are also their velocities $V_1$ and $V_2$. It follows that, when the two flying clocks rejoin in $O$, all clocks $O$, $O_1$ and $O_2$ agree on the fact that the clocks $O_1$ and $O_2$ have lost their original synchronization. In detail:
\begin{enumerate}
  \item According to $O$, the duration of round trip of $O_1$ is $\Delta t_1=2l_1/V_1$ and that of $O_2$ is $\Delta t_2=2l_2/V_2$
  \item The round trip proper duration of  clock $O_1$ will be $\Delta \tau_1= \Delta t_1\sqrt{1-V_1^2/c^2}$
  \item The round trip proper duration of  clock $O_2$ will be $\Delta \tau_2= \Delta t_2\sqrt{1-V_2^2/c^2}$
\end{enumerate}
It follows that the ratio between the two proper durations is:
\begin{equation}\label{rapportoterzetto}
\frac{\Delta
\tau_1}{\Delta\tau_2}=\frac{\Delta t_1}{\Delta t_2}\sqrt{\frac{1-V_1^2/c^2}{1-V_2^2/c^2}}=\frac{l_1}{l_2}\frac{V_2}{V_1}\sqrt{\frac{1-V_1^2/c^2}{1-V_2^2/c^2}}.
\end{equation}
\label{triplo}
If:
\begin{equation}\label{l1l2}
    \frac{l_1}{V_1}=\frac{l_2}{V_2},
\end{equation}
Eq. (\ref{rapportoterzetto}) assumes the simplified form:
\begin{equation}\label{rapportoterzetto2}
\frac{\Delta
\tau_1}{\Delta\tau_2}=\sqrt{\frac{1-V_1^2/c^2}{1-V_2^2/c^2}}.
\end{equation}
When the condition (\ref{l1l2}) applies, the durations of the round trips of the two flying clocks are the same in the reference frame of $O$.
In section \ref{hk}, we shall see that the asymmetric triplet journey effect -- under the condition (\ref{l1l2}) -- is the rectilinear version of Hafele\&Keating's experiment without a gravitational field.
\subsection{Experiments on time dilation and the journey effect}\label{expsec}
\subsubsection{Time dilation}
Nowadays, time dilation is analyzed through the Doppler effect.
 The time dilation factor, denoted as \(\gamma\), appears in the Doppler formula for photons emitted by an atom or nucleus moving with velocity \(v_1\) (see below Eq. (\ref{fredopplernu})).

However, this equation cannot be used directly to verify the time dilation effect. This is because the first-order Doppler term, which depends on \(v_1/c\), would overshadow the much smaller term dependent on \(v_1^2/c^2\). To address this, it is preferable to measure the light emitted in a direction perpendicular to the motion of the flying atoms. In this case, with \(\theta_1 = \pi/2\), the term dependent on \(v_1/c\) in Eq. (\ref{fredopplernu}) cancels out.

As a result, experimenters implemented specific configurations to eliminate the first-order Doppler effect. One widely used experimental setup was that introduced by Ives and Stilwell in 1938 \cite{ives}. Through the Doppler effect, the \(\gamma_1\) factor has been measured with an accuracy of \(10^{-9}\) \cite{Li}.
\par
Ives and Stilwell's experiment preceded the work of Rossi and Hall, who studied unstable particles. Rossi and Hall measured the lifetime of negative muons (then referred to as mesotrons) found in cosmic rays.

Starting in the 1950s, physicists began measuring the lifetimes of unstable particles produced by accelerators. A comprehensive review of the data collected can be found in a paper by Bailey et al.  \cite{bailey}. In experiments like that of Rossi and Hall, the average lifetime of particles in flight was measured indirectly by assessing their speed and the distance they traveled.

A key difference that significantly affected the accuracy and precision of these measurements was the transition from conducting experiments in the atmosphere to doing so in a vacuum. The accuracy achievable when studying the decay of unstable particles in a vacuum is much lower (approximately \(10^{-3}\)) compared to that achieved using the Doppler effect.
\par
Probably, students are told about the Rutherford's formula of radioactive decay:
\begin{equation}\label{decadimento}
N=N_0e^{-t/\tau},
\end{equation}
where $\tau$ is the decaying particles' (average) lifetime.
If the particles are in inertial motion with velocity $v$ in the laboratory IRS, we predict that they will cover an average distance before decaying given by:
\begin{equation}\label{distanza}
l= v\tau=v\frac{\tau_0}{\sqrt{1-v^2/c^2}},
\end{equation}
where $\tau_0$ is the particles' lifetime when at rest in the laboratory. The above formula has been experimentally corroborated.
\par
The experiment performed by Bailey et al. in 1977 \cite{bailey} was particularly intriguing.
They studied
the decay of negative or positive muons in uniform circular motion:
the result is identical to that which would have been obtained if the muons had been
in uniform rectilinear motion. Therefore, the authors conclude that
the acceleration to which muons are subjected ($a_c \approx
10^{18} {\rm g}$) does not affect the average muon lifetime. The authors also held that their experiment was a particular case of what we have called the journey effect. They wrote: ``The muons perform a round trip and so when compared with a
muon decaying at rest in the laboratory, simulate closely the so-called twin paradox, which was already discussed in Einstein's
first paper'' \cite[p. 304]{bailey}.
\subsubsection{The round-trip journey effect}\label{hk}
\begin{figure}[h]
\centering
\includegraphics[width=5cm]{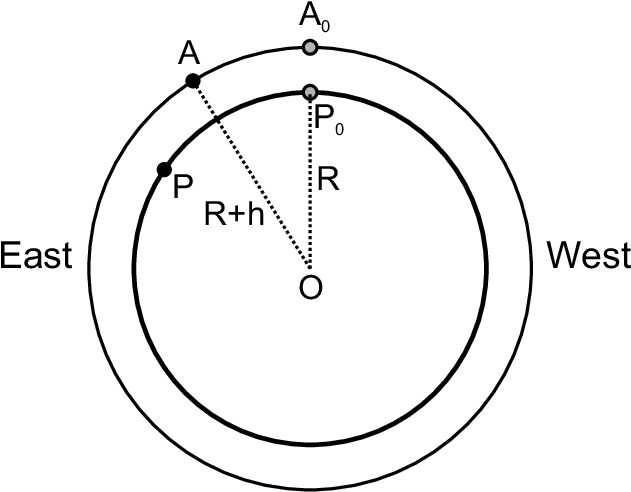}\\
\caption{Equatorial circumference seen by the North Pole.
The plane $A$ flights eastward with ground speed $v$. The ascent and descent parts of the plane's journey are neglected. The equatorial point  $P$ represents the departure/arrival airport. $P_0$ and $A_0$ are the airport's and plane's position at the time $t=0$. {\em In the non-rotating reference frame with origin at the Earth center $O$}, the airport $P$ moves eastwards with velocity $\Omega R$ and the plane with velocity $\Omega (R+h)+v$ (Galilean composition of velocities because both velocities are much smaller than $c$). Since $v < \Omega R$, the plane chases the airport $P$. In the case of the westward flight, the airport and the plane run towards each other. }\label{hafelefig}
\end{figure}
\par\noindent
It is important to note that the experiments conducted by Hafele and Keating, along with the subsequent experiment by Alley et al. \cite{alley}, are the only investigations directly examining the journey effect using macroscopic atomic clocks. Hafele and Keating utilized commercial aircraft to circumnavigate the Earth, while Alley and his team employed a dedicated aircraft that flew along a closed circuit.

In the theoretical analysis of Hafele and Keating's experiment \cite{hafeleajp}, one clock remains stationary on the Earth's surface at a point \( P \) on the equator. This clock is synchronized with another clock before the experiment begins. The second clock then ascends to a certain altitude and flies eastward along an equatorial path, returning to its original starting point (see Fig. \ref{hafelefig}). In a separate experiment, a clock on a plane circumnavigates the Earth in a westward direction. The ascent and descent phases of the flight are not included in the calculations.
\par
The flights are
described in a non-rotating reference system with the origin at the Earth's
center, which is in free fall (ECI - Earth-Centered Inertial). Therefore, the reference system chosen is
inertial. In this reference system, both clocks,
the one placed at the airport and the other in flight on the plane, are in
motion and immersed in the gravitational field
generated by the Earth's mass, assumed to be spherical.
\par
Hafele gave the theoretical treatment of the experiment
using the formalism of general relativity in the limit of weak gravitational fields and low velocities. The same
results can be obtained within special relativity using time dilation {\em and} the dependence of the fundamental period of an atomic clock on the gravitational
potential (see Appendix \ref{redshift}).
Moreover, we exploit the equivalence of the Hafele\&Keating experiment without a gravitational field with the asymmetric triplet journey effect. In the case of the eastward flight, the correspondence of the clocks is shown in the table \ref{eastward}.
\begin{table}
  \centering{
  \begin{tabular}{|l|c|c|c|}
    \hline
    Triplet & $O$& $O_1$ & $O_2$ \\
    H\&K & ECI& Airport & Airplane in eastward flight \\
    \hline
  \end{tabular}
  }
  \caption{Clocks' correspondence when comparing the asymmetric triplet journey effect with the Hafele and Keating experiment in the case of the eastward flight. In the ECI reference frame, the ``clock'' is a network of synchronized clocks. }\label{eastward}
\end{table}
\par\noindent
In the asymmetrical triplet journey effect, the motions occur in opposite directions, while in the eastward flight, the motions of the airport and the plane occur in the same direction. This difference is irrelevant because of the isotropy of space.
Notice that -- in the ECI reference frame -- the airport and plane's motion durations are the same (the ECI reference frame corresponds to the laboratory frame $O$ in the triplet journey effect).
It follows that, on the basis of Eq. (\ref{rapportoterzetto2}), we can write directly:
\begin{equation}\label{propereasstri}
\frac{\Delta
\tau_{East}}{\Delta\tau_{airport}}=\sqrt{\frac{1-u_{East}^2/c^2}{1-u_{airport}^2/c^2}}\approx \frac{1-u_{East}^2/2c^2}{1-\Omega^2R^2/2c^2},
\end{equation}
where $u_{East}=\Omega(R+h)+v$ and $u_{airport}=\Omega R$ are the velocity of the plane and the airport in the ECI reference frame. In the expression of $u_{East}$, we have used the Galilean composition of velocities because the velocities involved are $\ll c$.
First, we calculate the square of $u_{East}$ and rearrange its terms:
\begin{equation}\label{square?}
    u_{East}^2=\Omega ^2R^2\left( 1+\frac{h^2}{R^2}+\frac{2h}{R} +\frac{v^2}{\Omega^2R^2}+\frac{2v}{\Omega R}+\frac{2v}{\Omega R}\frac{h}{R}\right).
\end{equation}
Since we can neglect the terms in $h^2/R^2$ and in $h/R$, the equation (\ref{propereasstri}) becomes:
\begin{equation}\label{finaleeast}
    \frac{\Delta
\tau_{East}}{\Delta\tau_{airport}}\approx \frac{1-(\Omega^2R^2+v^2+2v\Omega R)/(2c^2)}{1-\Omega^2R^2/(2c^2)}\approx
1-\frac{v^2+2v\Omega R}{2c^2},
\end{equation}
where, in the last passage, we have neglected the term $\Omega^2R^2/2c^2$ compared to unity.
Notice that the dime dilation contribution to the mismatch between the two clocks (airport and plane) does not depend, in the approximation used, on the parameter $h$, i.e., on the height of the plane's route. Instead, this height comes in when considering the dependence of the clocks' fundamental period on gravity.
For weak gravitational fields, we have (Eq. (\ref{periodi}) of the Appendix \ref{redshift}):
\begin{equation}\label{fP}
T (R+h) \approx T  (R)\left(1 - {{gh}\over {c^2}} \right),
\end{equation}
namely, the fundamental period of the airplane's clock is smaller than that of the airport clock by a factor $(1-gh/c^2)$.
Hence, the ratio between the numbers shown by the two clocks in Eq. (\ref{finaleeast}) must be increased by a factor $1/(1-gh/c^2)\approx 1+gh/c^2$.
Therefore, the final equation is:
\begin{equation}\label{offset}
\frac{\Delta\tau_{East}^*}{\Delta \tau_{airport}^*}\approx 1+ \frac{gh}{c^2}- \frac{(2\Omega R+v)v}{2c^2},
\end{equation}
 that coincides with Eq. (8) of the Hafele's paper \cite{hafeleajp}.
\par
Before considering the westward flight, notice that: a) in the ECI reference frame, the durations of the plane's eastward and westward flights are equal; and b) in both eastward and westward cases, the durations of the airport's and plane's journeys are the same. Then, for westward flight, it is sufficient to change the sign of the velocity $v$ in Eq. (\ref{offset}):
\begin{equation}\label{offsetW}
\frac{\Delta\tau_{West}^*}{\Delta \tau_{airport}^*}\approx 1+ \frac{gh}{c^2}+ \frac{(2\Omega R-v)v}{2c^2}.
\end{equation}
The equations (\ref{offset}) and (\ref{offsetW}) show that there is a directional East-West effect due to the different relative velocities between the airport and the airplane in the two directions.
\par
The analysis of the experimental data collected is rather complex \cite{hafeleexp}. First,
it must be considered that the aircraft used did not travel
equatorial routes. Hence, the formula (9) of Hafele's work \cite{hafeleajp} must be
amended to take account of latitude and the component of
the aircraft's speed in the East/West direction. Moreover, the analysis was complicated because the aircraft did not maintain constant altitude, latitude, and ground speed. However, Hafele and Keating concluded that the expected effects
have been verified beyond reasonable doubt.
\par
Alley, who, a few years later, conducted a similar experiment with a group of
researchers from Maryland University,
commented: ``The comparison with the predictions seems to show an uncertainty of about 13\% for the westward direction, but much worse for the eastward direction. It is difficult to assign an uncertainty for the comparison of the individual potential and velocity effects, but their existence is certainly demonstrated \cite[p. 17]{alley}."
Alley's experiment, although characterized by superior accuracy
to that of Hafele and Keating (Alley estimated it around one percent
\cite[p. 22]{alley}), did not have the same resonance. On the other hand, in Alley's experiment, the East-West asymmetry was negligible.
Finally, It should be noticed that the theoretical treatment of the Hafele\&Keating experiment can be extended to satellites orbiting the Earth until the height $h$ of the orbit is sufficiently smaller than the Earth radius ($h\ll R$).
\par
The Hafele\&Keating experiment spectacularly attempted to corroborate the journey effect in a gravitational field using macroscopic clocks. However, its accuracy is very low when compared with that of the time dilation experiments with the luminous Doppler effect \cite{Li}, or Pound\&Rebka's experiment of the dependence of the fundamental period of a clock on the gravitational potential \cite{pound}.
\vskip2mm\par
\textsf{Implementing this section in a classroom relies more on the teaching context than other sections. However, at least some conceptual cornerstones should be discussed.}
\textsf{
\begin{enumerate}
\item The so-called ``clock paradox'' is not actually a paradox. It relates to the round-trip journey effect, which was first illustrated and discussed by Einstein in his 1905 paper \cite[p. 153]{ein05r}. In this work, Einstein refers to it as a ``peculiar consequence'' of his postulates. The term ``clock paradox'' was later coined and is extraneous to Einstein's original thoughts.
\item     The journey effect results from time dilation; however, it is important to note that time dilation and the journey effect are distinct phenomena. This distinction is further explored in the discussion regarding the triplet journey effect.
\item Teachers should cover some experimental validations of relativistic kinematic effects. Time dilation experiments involving unstable particles are commonly discussed in textbooks, often referencing Rossi and Hall's experiment. It is advisable to introduce experiments with atoms in flight in the classroom only after in-depth coverage of the Doppler effect for light.
\item Hafele and Keating's experiment is presented in some textbooks without delving into the calculations. Teachers could instead explain the simpler case of the asymmetric triplet journey effect and reason its conceptual equivalence to the Hafele \& Keating experiment conducted without a gravitational field.
\item With the widespread use of devices such as mobile phones and smartwatches that utilize the Global Positioning System (GPS), it is important for teachers to remember that these devices must account for time dilation and gravitational effects. Failing to do so can lead to significant errors in determining the device's position, making accurate location tracking impossible.
\end{enumerate}
}
The GPS provides experimental evidence for these relativistic effects. Students can be informed that when their mobile phone instructs them to ``Turn left in ten meters'', the measurement of ten meters is accurate because the necessary relativistic corrections have been applied. Without these corrections, the actual distance could diverge by as much as ten kilometers from what is indicated on the device. As Ashby puts it: ``If these relativistic
effects were not corrected for, satellite clock errors
building up in just one day would cause navigational
errors of more than 11 km, quickly rendering the system
useless \cite[p. 42]{gps}''.

\section{Dynamics}\label{dynamics}
\subsection{Doppler effect for photons}\label{dopp-pho-sec}
In modern high school curricula, light is commonly described in terms of light quanta. However, the Doppler effect is typically taught as a wave phenomenon. This approach is conceptually flawed, as it distorts the roles of these two descriptions of light and overlooks fundamental physical and epistemological questions regarding their relationship. Additionally, describing the Doppler effect in terms of photons presents an engaging application of relativistic dynamics based on the principles of energy and linear momentum conservation. Students are already familiar with elastic collisions, such as those involving billiard balls. In this context, the emission or absorption of a photon can be viewed as an intriguing inelastic collision occurring at the atomic scale.
\par
The first treatment of this topic was by Erwin Schr\"odinger \cite{erwin}. Schr\"odinger treated only the emission process and did not introduce into the calculations the energy difference $\Delta E$ between the two atomic energy levels involved. Schr\"odinger's paper was completely ignored. In 1938, Clinton Davisson published a similar paper, but he did not mention Schr\"odinger. This topic was resumed in 1990 by Red\u{z}i\'{c} \cite{redz}. More recent reevaluation of Schr\"odinger's paper and applications of energy and linear momentum conservation laws to interesting physical phenomena can be found in \cite{ggdoppler, rotanti, laser, natura}.
\begin{figure}[h]
\centerline{
\includegraphics[width=3cm]{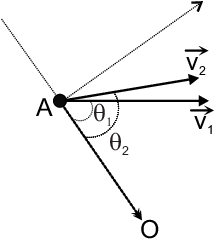}
}
\caption{\label{erwinfig}
Emission of a photon by an atom or nucleus $A$ in flight. The photon is emitted along the direction $A\rightarrow O$. The subscripts $1$ and $2$ refer to the quantities (velocities, angles) before and after the emission. The emission, which takes place in the plain determined by  $\vec v_1$ and ${AO}$, is assumed to be instantaneous and is described in the laboratory reference frame.}
\end{figure}
\par\noindent
\textsf{The complete relativistic treatment is too complex and lengthy to implement in a classroom setting. However, writing down the conservation equations governing the process can be instructive, particularly regarding the mass expressed as \(E/c^2\), where \(E\) is the rest energy of the atom. It's important to note that the rest energy differs for an atom in its ground state compared to when it is in an excited state; consequently, the mass of the atom changes accordingly.
}
\par
Energy conservation implies that:
\begin{equation}\label{energia2}
E_{ph}=\gamma_1 E_1^{emi}-\gamma_2 E_2^{emi}.
\end{equation}
$E_1^{emi}$ and $E_2^{emi}$ are the atom's rest energies before and after the emission. The two energies are related by the equation $E_1^{emi}-E_2^{emi}=\Delta E$, where $\Delta E$ is the energy difference between the two levels of the atomic transition. Precisely, we have: $E_1^{emi}= mc^2+\Delta E$ and $E_2^{emi}=mc^2$. Teachers should stress that $E_1^{emi}$, $E_2^{emi}$ e $\Delta E$ are relativistic invariant.
The conservation of the linear momentum along the direction of the photon emission and the perpendicular direction are (Fig. \ref{erwinfig}):
\begin{eqnarray}
\gamma_1 {{E_1^{emi}}\over{c^2}}v_1 \cos\theta_1 &= & \gamma_2 {{E_2^{emi}}\over{c^2}}v_2 \cos\theta_2 +{{E_{ph}^{emi}}\over{c}}
\label{qmx}\\
\gamma_1 {{E_1^{emi}}\over{c^2}}v_1 \sin\theta_1 &= & \gamma_2 {{E_2^{emi}}\over{c^2}}v_2 \sin\theta_2.\label{qmy}
\end{eqnarray}
It turns out that:
\begin{equation}\label{fredoppler}
E^{emi}_{ph}=E^{emi}_{{ph}_0}\frac{\sqrt{1-v_1^2/c^2}}{1-(v_1/c)\cos\theta_1}; \qquad E^{emi}_{{ph}_0}
=\Delta E\left(1- \frac{\Delta E}{2(mc^2+\Delta E)}  \right),
\end{equation}
where $v_1$ is the atom's velocity before the emission, $E^{emi}_{ph}$ is the energy of the emitted photon,  $E^{emi}_{{ph}_0}$
the energy of the photon emitted when the atom is at rest before the emission, and $\Delta E$ is the energy difference between the two energy levels.
For atoms, $\Delta E\ll mc^2$. Hence, $E^{emi}_{{ph}_0}= \Delta E(1- \Delta E/2mc^2)$.
Dividing both members of Eq. (\ref{fredoppler}) by $h$, we obtain:
\begin{equation}\label{fredopplernu}
\nu=\nu_0\frac{\sqrt{1-v^2_1/c^2}}{1-(v_1/c)\cos\theta_1}.
\end{equation}
\textsf{The formula mentioned above aligns with the results obtained from the wave description of light. Teachers should take this opportunity to discuss the different physical meanings of equations (\ref{fredoppler}) and (\ref{fredopplernu}), as well as the importance of having clear criteria for when and how to apply them.
The corpuscular description of light should be used when addressing the emission or absorption of light by atoms or nuclei, as initially suggested by Einstein \cite{ein05lq}. On the other hand, the wave description can be applied when the number of photons involved -- whether emitted one at a time or all together -- is statistically significant.}

\textsf{The theoretical analysis of the double-slit experiment, conducted with one photon at a time \cite{single}, is particularly enlightening. In this case, we must utilize the corpuscular description (in terms of probability) to describe the individual spot on the detector. However, when a sufficient number of photons have been deposited on the detector, we can adopt the wave description \cite{natura}. }

\textsf{It turns out that the corpuscular and wave descriptions of interference share the same mathematical structure. The transition from a wave description to a corpuscular one is based on converting the wave intensity at a point on the detector to the probability of detecting a photon at that same point. A similar conceptual framework was illustrated by experiments involving images taken with varying numbers of photons. These observations indicate that speaking of wave-particle dualism is not meaningful; instead, we have two different descriptions and know when to use each.}

\textsf{However, after presenting this framework, teachers will likely face a common question from students: ``What is light?'' This question cannot be answered without an epistemological reflection based on two fundamental principles: a) Physical theories aim to predict the values of physical quantities, but they do not describe ``what is happening in the world''. b) An ontological assertion regarding the existence of a theoretical entity in the world can only be made a posteriori, must align with acquired knowledge, and is only plausible. If this plausibility holds, then the ontological assertion can be considered probable. For a more in-depth discussion of this issue, see the references in \cite[sec. 8]{natura} and \cite[chapter I]{erq}.
}
\par
In the case of absorption, after having re-written adequately the conservation equations, we get:
\begin{equation}
E_{ph}^{abs} = E^{abs}_{{ph}_0} {{\sqrt{1-v_1^2/c^2}}\over{1 -(v_1/c)\cos \theta_1 }};\qquad E^{abs}_{{ph}_0}=\Delta E \left( 1 + \frac{\Delta E}{2mc^2}\right). \label{ugualeass}
\end{equation}
Notice that the equations for emission (\ref{fredoppler}) and absorption (\ref{ugualeass}) differ only in the value of $E_{ph_0}$. Finally, it can be shown that the photon energy can also be written in terms of the quantities (velocities and angles) after the emission/absorption (\cite[p. 27]{natura}). This result is not surprising because, given these quantities before emission/absorption, the exact quantities after emission/absorption are determined by the conservation equations.
\par
\textsf{ A Newtonian approximation for low velocities of the emitting or absorbing atom can provide valuable insights. Teachers should emphasize that while the wave description relates the wave frequencies in two reference systems (IRS), the corpuscular description offers a complete quantitative understanding of the emission or absorption of a light quantum by an atom. As Einstein stated: ``[\dots] it is conceivable that despite the complete confirmation of the theories of diffraction,
reflection, refraction, dispersion, etc, by experiment, the theory of light, which
operates with continuous spatial functions, may lead to contradictions with experience when it
is applied to the phenomena of production and transformation of light'' \cite[p. 87]{ein05lq}.}
\par
\textsf{An atom of mass $m$ flights with velocity $v_1\ll c$ along the positive direction of the $x$ axis. If the atom absorbs a photon of energy $E_{ph}$  in a head-on collision (the photon is flying along the negative direction of the $x$ axis),
the conservation of the linear momentum reads:
\begin{equation}\label{momentum}
mv_1-\frac{E_{ph}}{c}=mv_2,
\end{equation}
where $v_2$ is the atom's velocity after the absorption. Teachers should observe that photon linear momentum introduces a relativistic quantity into Newtonian dynamics.
The conservation of energy implies that:
\begin{equation}\label{energy}
\frac{1}{2}mv_1^2+ E_{ph}=\frac{1}{2}mv_2^2+\Delta E,
\end{equation}
$\Delta E$ is the energy difference between the two atomic levels. $\Delta E$ is a relativistic invariant, i.e., it has the same value in every IRS.
From Eq. (\ref{momentum}), we get:
\begin{equation}\label{velocity}
v_2=v_1-\frac{E_{ph}}{mc}
\end{equation}
Inserting this value into Eq. (\ref{energy}) we obtain:
\begin{equation}\label{energy2}
E_{ph}= \Delta E+\frac{1}{2mc^2}E_{ph}^2-B_1E_{ph},
\end{equation}
where $B_1=v_1/c$.
Notice that in Eq. (\ref{energy2}), a term $mc^2$ has appeared and has the dimensions of energy. We shall come back to this novelty later.
If we write:
\begin{equation}\label{position}
E_{ph}=\Delta E(1+\alpha),
\end{equation}
where $\alpha\ll 1$, Eq.
(\ref{energy2}) assumes the form:
\begin{equation}\label{energy3}
\Delta E(1+\alpha)=\Delta E +\frac{\Delta E}{2mc^2}\Delta E(1+\alpha)^2 -B_1\Delta E(1+\alpha).
\end{equation}
Defining the dimensionless, positive parameter:
\begin{equation}\label{B_T}
B_T=\frac{\Delta E}{2mc^2}\ll 1,
\end{equation}
we see that in Eq. (\ref{energy3}) all the dimensionless parameter $\alpha, B_1, B_T$ are much less than one.
To give some orders of magnitude. For the Hydrogen line $\lambda=1215.67 \times 10^{-10}\, {\rm m}$, we have $B_T=5.43\times 10^{-9}$. $B_1$ may be assumed $\gg B_T$ and $\le 10^{-3}$.
Hence, keeping in Eq. (\ref{energy3})  only the terms of the first order in $\alpha, B_1, B_T$, we get:
\begin{equation}\label{finale19}
\alpha= B_T-B_1.
\end{equation}
And, finally:
\begin{equation}\label{finale2}
E_{ph}=\Delta E(1+B_T-B_1).
\end{equation}
The variation of the atom's kinetic energy due to the absorption of the photon is given by:
\begin{equation}\label{cineticavar}
\Delta E_K= E_{ph}-\Delta E=\Delta E(B_T-B_1).
\end{equation}
If $B_1=0$, $E_{ph}=\Delta E(1+B_T)$ and $\Delta E_K=\Delta E^2/2mc^2=E_R$. $E_R$ is the atom's {\em recoil energy}.
}
\par
\textsf{Depending on the teaching context, teachers should find their path through the above calculations and discuss some (or all)  the following points.
The energy of the absorbed photon is always different from $\Delta E$, the energy difference between the two atomic energy levels, unless $B_1=B_T$ (see Eq. (\ref{finale2})). The parameter $B_T=\Delta E/2mc^2$ is a threshold parameter \cite{laser}.
In normal conditions, $B_T$ can be neglected if compared with $B_1$. Hence Eq. (\ref{finale2}) reduces to:
\begin{equation}\label{finale5}
E_{ph}=\Delta E(1-B_1),
\end{equation}
and Eq. (\ref{cineticavar}) assumes the form:
\begin{equation}
\Delta E_K= -B_1\Delta E.
\end{equation}
In a head-on collision, these two last equations show that the photon energy necessary for absorption is less than $\Delta E$ by an amount equal to $B_1 \Delta E$ and that this missing amount is supplied by a decrease of the atom's kinetic energy.
}
\par
\textsf{Naturally, teachers should rewrite Eq. (\ref{finale5}) in terms of frequencies, a quantity more familiar to students when studying the Doppler effect. Dividing both members of Eq.  (\ref{finale5}) by $h$, we get:
\begin{equation}\label{nunu}
\nu = \nu_0\left(1 -\frac{v_1}{c}  \right).
\end{equation}
}
\textsf{The recoil energy - i.e., the kinetic energy imparted to the atom by the absorption of a photon when the atom is at rest before absorption is given, as we saw above, by the equation:
\begin{equation}\label{rinculo}
E_R=\frac{\Delta E^2}{2mc^2}.
\end{equation}
The term $mc^2$ is extraneous to Newtonian dynamics. Special relativity tells us that it is the energy associated with the mass $m$. This foreign visitor came sneakily in through the photon linear momentum of Eq. (\ref{momentum}).
}
\subsection{${\rm E=mc^2}$}\label{mc2sec}
The young Enrico Fermi wrote in 1922 \cite{fermi_me}:
\begin{quote}\small
The immense conceptual importance of the theory of relativity, as a contribution to a deeper understanding of the relationship between space and time, and the lively -- often passionate -- discussions it has consequently sparked even outside strictly scientific circles, may have somewhat diverted attention from another of its results. Though less sensational and, let us admit it, less paradoxical, this result has no less notable consequences in physics and is likely to grow in significance as science progresses. The result we refer to is the discovery of the relationship linking a body's mass to its energy
  [\dots] It will rightly be said that there appears to be no possibility, at least in the near future, of finding a way to release these terrifying quantities of energy -- a prospect one can only hope remains unrealized -- as the explosion of such a colossal amount of energy would first and foremost blow to pieces the unfortunate physicist who managed to produce it.
\end{quote}
As we know, scientists and engineers freed the energy contained in nuclei, first with a controlled chain reaction (1942), then with the fission bomb (1945), and Fermi contributed to both breakthroughs.
\par
\textsf{In high school, it is not possible to derive two fundamental equations of relativistic dynamics (instead, it might be possible in elementary physics courses):
\begin{eqnarray}
\mathcal {E}_0 &=& mc^2\\\label{mc2}
\mathcal {E}  &=& \gamma mc^2; \qquad \gamma =\frac{1}{\sqrt{1-v^2/c^2}},\label{gammamc2}
\end{eqnarray}
where $\mathcal {E}_0$ is the rest energy of the mass $m$ and $\mathcal {E}$ the energy of the mass $m$ at velocity $v$.
However, it is possible to bring out the term $mc^2$, treating the absorption of a photon by an atom with Newtonian dynamics (see preceding section). As far as Eq. (\ref{gammamc2}),  it must be assumed: we can only show that, in the limit of $v\ll c$, it yields the Newtonian expression of the particle's kinetic energy $\mathcal {E} - \mathcal {E}_0 = (1/2)mv^2$.
}
 \begin{figure}[h]
\centering{
\includegraphics[width=7cm]{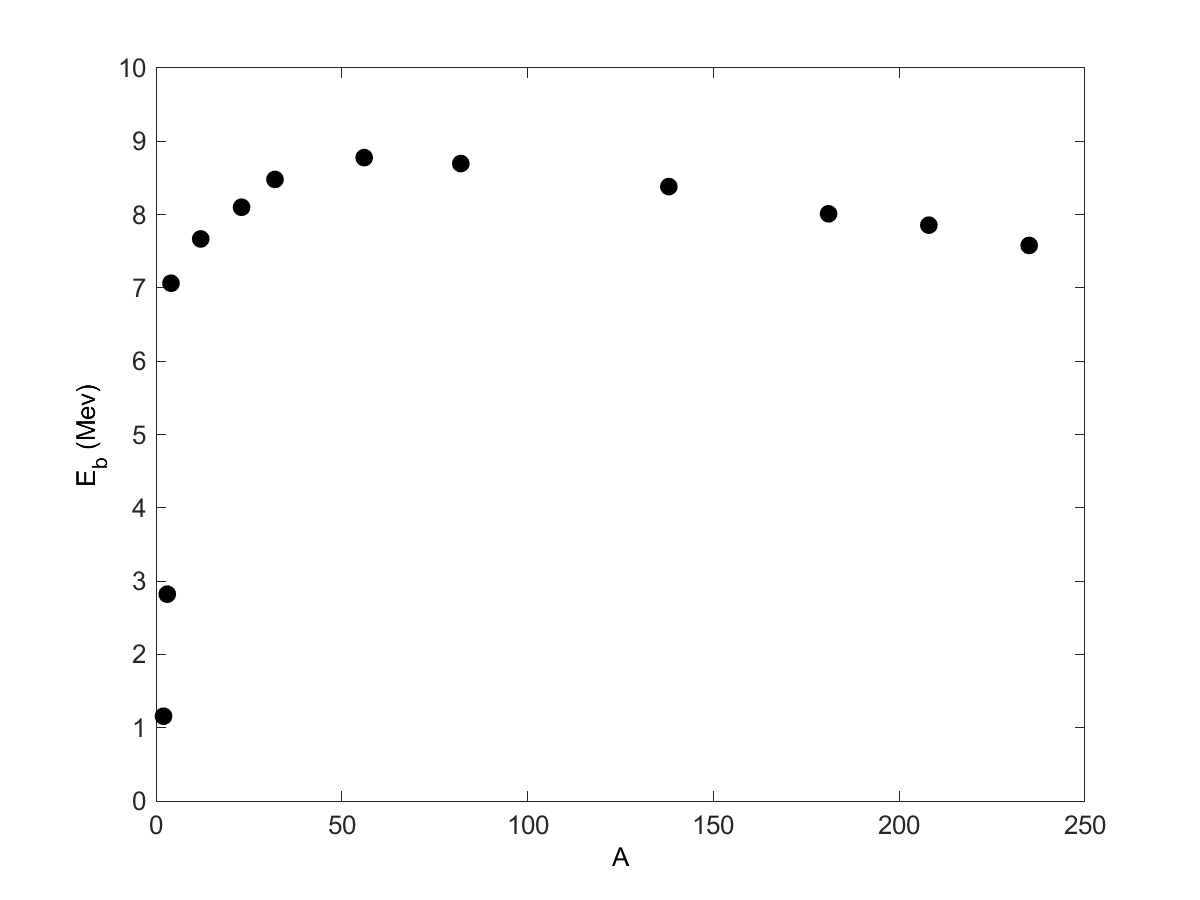}
}
\caption{Nucleon average binding energy as a function of the mass number $A$ (number of nucleons).}\label{lega}
\end{figure}
\par\noindent
\textsf{Teachers could develop the following considerations when returning to Eq. (\ref{mc2}).
If $n$ particles of mass $m_i; i=1,2\dots n$ are infinitely distant from each other and at rest in the laboratory reference frame,
the rest energy of the system made of all particles is $\sum_i m_ic^2$. Now, if the $n$ particles join together  to form a composite particle,
the rest energy  $E_0$ will be given by:
\begin{equation}\label{difetto-massa}
E_0=\sum\nolimits_{i}^{}m_ic^2+E_{binding}=Mc^2,
\end{equation}
where $E_{binding}$ is the binding energy of the composite particle, and  $M$ is its mass.
Since in a bound state $E_{binding}<0$, the mass of a composite particle is more petite than the sum of the masses of its constituents (mass defect). This result could be illustrated by considering the formation of a hydrogen atom from its constituents, a proton and an electron. During the capture of an electron by a proton, one or more photons will be emitted because the electron goes from an unbound (free) state to the ground state (Bohr's model). The equation of energy conservation for this process reads:
\begin{equation}\label{hclassica}
E_{binding} =- E_{rad},
\end{equation}
$E_{rad}>0$ being the energy of the radiation emitted during the formation of the atom. It turns out that the mass of the hydrogen atom is smaller than the sum of the masses of its constituents.
 Naturally, for the hydrogen atom, the mass defect is negligible. Moreover, to free the electron of the hydrogen atom, we must supply an energy equal to or greater than the ionization energy, and in this process, there is no energy gain.
}
\par
\textsf{In the case of nuclei, the situation is different. If a nucleus of mass $M$ contains $N_p$ protons and $N_n$ neutrons, the average binding energy $E_b$ for a nucleon (proton or neutron) is defined as (with obvious notations):
\begin{equation}\label{bond}
E_b=-{{[M - (N_p m_p+N_n m_n  )]}\over{N_p+N_n}}\,c^2.
\end{equation}
Hence, the energy freed during a fission or fusion reaction is given by:
\begin{equation}\label{fission}
E=N(E_b^f-E_b^i),
\end{equation}
where $N$ is the nucleon number involved in the reaction and $E_b^i$, $E_b^f$ is the average nucleon binding energy in the initial and final state, respectively.
As shown in Fig. \ref{lega}, the average binding energy per nucleon increases until about $8.7\, {\rm MeV}$ in correspondence with the mass number $57$ and gradually decreases. Hence, the fusion of two light nuclei or the fission of a heavy nucleus, adequately chosen, will free energy as the particles' kinetic energy or photons. However, there is a profound difference between fission and fusion processes. Indeed, a fission process is triggered by a slow (thermal) neutron produced by a radioactive process so that the energy freed by the fission of a nucleus is, practically,  all that predicted by Eq. (\ref{fission}). Instead, we must force nuclei to fuse at the expense of a significant amount of energy, thus making the net balance very critical.
}
\section{Discussion}\label{discussion}
 We can derive the transformations for the coordinates \((x, y, z, t)\) by assuming the homogeneity and isotropy of space, the homogeneity of time, and a kinematic principle of relativity. The resulting transformations resemble Lorentz's, but they replace the constant \(c\) with an unspecified velocity limit \(\kappa\). The choice of \(\kappa = c\) is necessitated by the independent assumption regarding the validity of Maxwell's equations in a vacuum. In Italy, the reference work is that of Carlo Cattaneo (1958) \cite{cattaneo}. Cattaneo's paper, published in the {\em Rendiconti dell'Accademia dei Lincei}, was largely overlooked, likely due to the language used and the limited reach of the journal. In \cite[pp. 42-45; 80-89]{ggib}, Cattaneo's derivation was revisited under the principle of causality. A more recent paper proposed a derivation similar to Cattaneo's but was unaware of his work \cite{copia}.

Cattaneo's derivation results in the following coordinate transformations:
\begin{eqnarray}\label{def}
x' & = & {{1} \over {\sqrt {1+V^2/\alpha}}}\, (x-Vt)\nonumber\\
&&\nonumber\\
y' & = &  y  \nonumber\\
&&\\
z' & = &  z  \nonumber\\
&&\nonumber\\
t' & = & {{1} \over {\sqrt {1+V^2/\alpha}}} \left(t+ {{V}\over{\alpha}}x \right),\nonumber
\end{eqnarray}
where $\alpha\neq 0$.
The option $\alpha>0$ is ruled out by considering the following thought experiment.
In the IRS $K'$ a signal (unspecified) is sent from $x_1'$ at the time $t_1'$ along the positive direction of the $x'\equiv x$ axis. Suppose that the signal reaches $x_2'$ at the time $t_2'$.
In the IRS $K$ the interval of time between the two events, departure and arrival of the signal, is given by:
\begin{equation}
  \Delta t={{1}\over{\sqrt {1+V^2/\alpha}}} \left(\Delta t'-
{{V}\over{\alpha}}\Delta x'
  \right).
\end{equation}
If $\alpha>0$, when
\begin{equation}
 {{\Delta x'}\over{\Delta t'}}=\delta'>{{\alpha}\over{V}} ,
\end{equation}
we get that $\Delta t<0$ ($\delta'=\Delta x'/ \Delta t'$  is the signal's velocity in $K'$).
Consequently, in $K$, the signal arrival precedes its departure, thus violating the causal principle.
\par
Instead, if $\alpha<0$, putting $\alpha =-\kappa ^2$, equations (\ref{def}) assume the form:
\begin{eqnarray}\label{def_kappa}
x' & = & {{1} \over {\sqrt {1-V^2/\kappa^2}}}\, (x-Vt)\nonumber\\
&&\nonumber\\
y' & = &  y  \nonumber \\
&&\\
z' & = &  z  \nonumber \\
&&\nonumber\\
t' & = & {{1} \over {\sqrt {1-V^2/\kappa^2}}} \left(t- {{V}\over{\kappa^2}}x \right).\nonumber
\end{eqnarray}
Developing Cattaneo derivation, a rule for the velocities composition along the $x\equiv x'$ axis is obtained at a certain point of the calculation. This rule, in the case of $\alpha=-\kappa ^2$, assumes the familiar form:
\begin{equation}\label{velox2}
u_x'= {{u_x- V} \over {1 -u_xV/\kappa ^2}},
\end{equation}
where $\kappa$ is, at a glance, a velocity limit.
\par
\textsf{Lorentz transformations are fundamental to special relativity, and the speed limit $(c)$ is essential to Lorentz transformations.
In elementary physics courses or at high school, teachers have two choices: to simply write Lorentz transformations or to deduce them as shown in section \ref{LT} using the exchange of light flashes between two IRS. Indeed, a treatment like that of Cattaneo is too long and formal to be proposed in elementary physics courses or at high school. However, the presentation to the students of Cattaneo's derivation, as sketched above, should have a rewarding conceptual impact.
}
\par
In principle, special relativity could be taught using the space-time concept. The reference book is that of Taylor and Wheeler \cite{T&W}. De Ambrosis and Levrini discussed the features of this kind of approach with in services teachers engaged in a Master course on the teaching of modern physics \cite{anna-olivia}. The use of Minkowski's diagrams in upper high school has been tested with positive feedbacks \cite{argentina}. Recently, the space-time approach was based on the use of ``mechanical instrument that allows to experience Special Relativity hands-on'' \cite{leonardi}. A careful, valuable ``guide for teachers'', based on the space-time approach, was proposed some years ago by Elio Fabri \cite{fabri}. However, no in-class controlled trials on a general space-time approach like that proposed by Taylor and Wheeler or by Fabri have been carried out to our knowledge.
\par
As shown in the preceding sections, thought experiments with the exchange of light flashes between two IRS allow us to derive all the kinematics effects of special relativity and the Lorentz transformations using simple equations. Within this approach, a central role is played by the propagation of light in a vacuum (speed limit) and by the quantum-relativistic interaction between light and matter (emission and absorption of light quanta by atoms or nuclei). The emphasis on light as a relativistic phenomenon is a signature of this teaching approach.
\par
Independently of the approach chosen, implementing a new teaching proposal in the classroom is a laborious process requiring several conditions: the teacher's adhesion to the conceptual framework of the proposal, the creation of a teaching module, experimentation in class, and the use of agreed-upon tools to evaluate the student's learning. Therefore, much work remains. Meanwhile, we would like to briefly report on two field tests.
\par
The few hours available limited the duration of the tests. As often happens, the class profile was inhomogeneous, presenting a challenge many educators can relate to. Class discussions, heterogeneous groups, and homework addressed this diversity.
\par
In the first test [M. G. B., ten hours], essential elements of special relativity were presented along with Newtonian kinematics with an explicit warning to students that they would deepen the issue in the following years of study.
Simulations produced using the GeoGebra software were a valid aid to the students' visualization of the proposed work: inertial reference frames in relative motion and the exchange of flashes of light between two inertial reference frames. After the simulations were reproduced and discussed, the students were invited to manipulate and modify them. This step was essential to mastering the physics behind the simulations.
\par
The postulate that the value of light speed in a vacuum is the same in all inertial reference frames was preceded by the vision of a video on the historical development of hypotheses and measurements of light speed. The Lorentz transformations were given without proof, and the Galilean ones derived from the former in the limit $c\rightarrow \infty$. The passage was proposed as an approximation of the calculation that neglects very small addends, emphasizing that Galilean transformations can be used in cases where the relative velocities of the inertial reference frames are minimal compared to light. The composition of velocities used in exercises was the Galilean one; however, it was compared with the relativistic formula.
\par
To address the study of time dilation, a thought experiment involving the exchange of flashes of light between two inertial reference frames, introduced by an exemplary simulation (\href{https://www.geogebra.org/m/mm2uczpm}{here}), was proposed. The simulation was based on the calculations developed in sections \ref{k-method} and \ref{TDsec}.
The time dilation formula was derived by stimulating the contribution of students who worked collaboratively. This part of the work lasted about three hours.
Always preceded by a simulation (\href{https://www.geogebra.org/m/wv3tz3yk}{here}), the study of the relativity of simultaneity, based on section \ref{train} above, was dealt with as a problem posed to the students.
\par
This trial suggests that the approach to relativistic kinematics effects by exchanging flashes of light of null duration between two inertial reference frames is suitable. Students have proven they can master equations like $l=l_0+v\Delta t$ to derive the fundamental time dilation formula.
\par
 The second trial [M. L., eleven hours] was composed of two units and tested in a class of the fourth year of a Scientific Lyceum (2024) \footnote{The teaching module was conceived for the third year. However, the teaching context made it possible only in a fourth-year class.}.The first aimed to improve students' ability to graphically represent the uniform motion of a particle or a flash of light while studying Newtonian kinematics. Initially, the students were asked to draw a graphic of the inertial motion of a particle using, as usual, a $(t,x)$ plane, then using a $(x, t)$ plane. The next step was using a plane $(x, ct)$, where both coordinates have the dimensions of a length. Naturally, in this plane, the motion of a light flash along the $x$ axis is represented by the bisector of the first quadrant. Instead, the bisector of the second quadrant represents the motion of a light flash along the negative direction.
In Newtonian kinematics, all particle velocities are allowed. However, students were told to limit their drawing of particles' motion to velocities less than that of light because they would soon apprehend that the light velocity in a vacuum is a speed limit. The exercises proposed to the students had the goal of accustoming them to using a $(x,ct)$ plane given its exploitation in the graphical representation of thought experiments like those on the relativity of simultaneity discussed in section \ref{train} and in Appendix \ref{fulmini}.
\par
The second item concerned the distinction between inertial and non-inertial reference systems. As is typical in teaching practices, the usual definition of an inertial reference frame is followed by a discussion of the approximations under which a laboratory on the Earth's surface can be considered an inertial reference frame. The disturbing but unavoidable presence of the Earth's gravitational field was exploited to find an operative method to distinguish between inertial and non-inertial reference frames.
\par
With his arms extended horizontally, the teacher (or a student) holds two balls (not magnetic or electrically charged). If the teacher lets the balls free, they fall vertically. What happens if the teacher repeats the experiment while moving in a rectilinear uniform motion? And, again, what happens if the teacher's motion is uniformly accelerated? Very qualitative experiments were performed in class. The unit was completed with the presentation and discussion of a simulation in which the balls' fall was viewed from the classroom's reference system. A second simulation (not presented to the students) is possible and enjoyable. This simulation describes the thought experiment in the moving reference frame $K'$. In $K'$, while uniformly accelerated along the positive direction of the $x\equiv x'$ axis, the balls are subjected to two acceleration fields: the gravitational field $g$ and the pseudo-gravitational field $-a$. Hence, the balls' motion is the composition of two perpendicular uniformly accelerated motions. Their trajectories are straight lines of the type:
 \begin{equation}\label{retta}
    y'=y'_0+\frac{g}{a}(x'-x'_0),
 \end{equation}
where $y'_0$ is the balls' initial height and $a$ is the acceleration of  $K'$. By measuring $y'_0$ and $x'$ corresponding to $y'=0$, $K'$ measures the proper acceleration. \href{https://www.geogebra.org/m/rrn2ypm6}{Here}, the reader will find the two simulations composed in a single one.
\par
The next step is to propose some teaching modules based on the conceptual network discussed in the preceding sections. This step is within our reach and will be accomplished as a proposal covering the last three years of a Scientific Lyceum.
Implementing controlled experimentation in an adequate number of classes will be much more challenging.
\section{Conclusions}
As the available literature suggests, students' comprehension of the fundamental principles of special relativity is often hindered by preconceived notions, familiarity with Galilean and Newtonian physics, and discrepancies between some predictions of special relativity and everyday experiences.

This paper aims to provide teachers in elementary physics and high school courses with a tool to teach the essential features of special relativity while considering the challenges that students face, as highlighted by various studies. The choice of how and when to use this tool is left to the teachers, who can select their preferred approach from several options.

 Our proposal presents special relativity as a solution to the difficulties encountered in Newtonian dynamics, particularly illustrated by the infinities that arise in Newtonian uniformly accelerated motion.
The core element of our proposal involves using thought experiments that depict the exchange of light flashes of null duration between two inertial reference frames in relative motion. This method helps derive the kinematic effects of special relativity. As a further support, we include visual representations of $(x, ct)$ planes and simulations that illustrate the results of simple calculations.
\par
Additionally, we discuss experimental corroborations of the kinematic effects of special relativity, complementing the theoretical analysis. Among them, we analyze the Hafele and Keating's experiment, which is explained solely using special relativity. Furthermore, the Doppler effect, typically addressed within the wave description of light, is explored here as an application of energy and linear momentum conservation during the emission or absorption of a photon by an atom (or nucleus). The solution to this problem within Newtonian dynamics leads to the emergence of the term $mc^2$ in the expression for the recoil energy of the atom. This surprising result stems from incorporating the linear momentum of the photon, $E_{ph}/c$, into the conservation equation for linear momentum.

Lastly, preliminary classroom tests conducted by two authors indicate the need for a broader study, including standardized evaluations of students' learning.

\begin{appendices}
\small
\counterwithin*{equation}{section}
\renewcommand\theequation{\thesection\arabic{equation}}
\section{Length contraction with light flashes }\label{LC}
Consider the arrangement
of the Fig. \ref{piattaforma}.
As we shall see, $O'$ can measure both the velocity and the length of the stick $OB$ (this derivation strictly follows the one of \cite[pp. 260-262]{erq}).
\begin{figure}[h]
\centering{
\includegraphics[width=5cm]{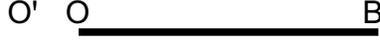}\\
}
\caption{The clock $O'$ and the stick $OB$ approach in relative, head-on inertial motion with relative velocity $V$.   Both reference frames consider the relative velocity $V$ as positive in their calculations.}\label{piattaforma-app}
\end{figure}
\par\noindent
When $O'$ meets $O$ at the time $t_0'$, $O'$ emits towards $B$ a flash of light: the flash is reflected by
$B$ and reaches $O'$ at the time $t_1'$; it is reflected by $O'$, reaches $B$ and, after being reflected again by $B$, meets $O'$ at the time $t_2'$.
Notice that $O'$ makes three reads (Table \ref{3}).
\begin{table}[h]
\centering{
\begin{tabular}{|c|l|}
\hline
Time & Definition\\
$t_0'$ & $O'$ meets $O$ and emits a flash towards $B$\\
& \\
$t_1'$ &  $O'$ receives the flash reflected by $B$\\
&\\
$t_2'$& $O'$ receives the flash reflected by $B$ the second time\\
&\\
\hline
\end{tabular}
}
\caption{The three times read by the clock $O'$.}\label{3}
\end{table}
\par\noindent
The distance $d'$ at which $B$ receives and reflects the flash for the first time is given by:
\begin{equation}\label{primo}
d'= \frac{c}{2} (t_1'-t_0').
\end{equation}
On the other hand, if $t'_R=t'_0+(t'_1-t'_0)/2$ is the time at which $B$ reflects the first flash, the distance $d'$ can be written as:
\begin{equation}\label{altra}
d'= l'- V(t'_R-t'_0)=l'-\frac{V}{2}(t_1'-t_0'),
\end{equation}
where $l'$ is the stick's length in the reference frame of $O'$.
Hence, from Eqs. (\ref{primo}) and (\ref{altra}), we get:
\begin{equation}\label{primo_t}
t_1'-t_0' = {{2l'}\over{c+V}}.
\end{equation}
With the same procedure, we get, for the time interval $t_2'-t_1'$;
\begin{equation}
t_2'-t_1'  =  {{2[l'-V(t_1'-t_0')]}\over{ c+V}}.
\end{equation}
Hence, taking into account Eq. (\ref{primo_t}):
\begin{equation} \label{secondo_t}
t_2'-t_1'  = 2l' {{c-V}\over{(c+V)^2}}.
\end{equation}
Putting:
\begin{eqnarray}
t_1'-t_0'&=&\alpha>0\nonumber\\
t_2'-t_1'&=&\beta>0,\nonumber
\end{eqnarray}
we get from Eq. (\ref{primo_t}):
\begin{equation}\label{primo_t1}
2l'=\alpha (c+V),
\end{equation}
and, from Eq. (\ref{secondo_t}):
\begin{equation}
2l'=\beta{{(c+V)^2}\over{c-V}}.
\end{equation}
From these two equations, we obtain:
\begin{equation}
V=c\,{{\alpha-\beta}\over{\alpha+\beta}}
=c\,{{(t'_1-t'_0)- (t'_2-t'_1)}\over{t_2'-t_0'}}.
\end{equation}
Dividing member by member Eq. (\ref{primo_t}) and Eq. (\ref{secondo_t}) we find that $\alpha>\beta$. Therefore, $V>0$, as it was assumed.
Putting this value of $V$ into equation(\ref{primo_t1}), we get finally:
\begin{equation}\label{contratempi}
l'=c\,{{\alpha^2}\over{\alpha+\beta}}=c\, {{(t_1'-t_0')^2}\over{t_2'-t_0'}}.
\end{equation}
So, we have shown that $O'$ can deduce the value of $V$ and $l'$ by reading three numbers on its clock.
\par
For the relativity principle, $O$ describes the measurements made by $O'$ by writing down the same equations written by $O'$ in which, however, the primed quantities are substituted by the unprimed ones. For instance, $O$ writes:
\begin{equation}\label{primo_to}
t_1-t_0 = {{2l_0}\over{c+V}};\qquad t_2-t_1  = 2l_0 {{c-V}\over{(c+V)^2}},
\end{equation}
where $l_0$ is the proper length of the stick $OB$.
Hence, the Eq. (\ref{contratempi}) is substituted by the equation:
\begin{equation}\label{analoga}
l_0=c\, {{(t_1-t_0)^2}\over{t_2-t_0}}.
\end{equation}
While the measurements are performed by $O'$ at the same point $O'$, $O$ sees the measures made by $O'$ at different points of its reference system. Therefore, the relation between the duration of the same phenomenon, evaluated in the two references, is given by $\Delta t'=\Delta t/\Gamma$. Consequently, from Eqs. (\ref{contratempi})  and
(\ref{analoga}), we get immediately the equation of the ``length contraction":
\begin{equation}\label{con00}
l'=l_0\sqrt{1-\frac{V^2}{c^2}}.
\end{equation}
Like time dilation, the length contraction is a symmetrical phenomenon.
\textsf{The above calculation is lengthy. However, it bestows two conceptual rewards: a)  using the same method to measure the length of a stick at rest or in motion; b) showing that the `length contraction' is the result of a measurement made with light pulses, a method already familiar to students after studying time dilation and the Doppler effect. Teachers should add that a similar method is used by radar ranging in aviation, where, however, the contraction of the distance between the radar station and the airplane is  neglected owing to its low velocity
($V=10^3\, {\rm km\, hr^{-1}} \rightarrow V/c\approx 9.27\times 10^{-7}$).
}
\section{Einstein's second thought experiment on relativity of simultaneity}\label{fulmini}
\begin{figure}[h]
\centering{\includegraphics[width=6cm]{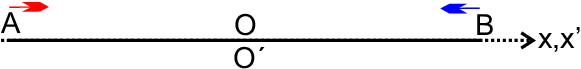}\\
}
\caption{\small {$O'$ is the center of a train  $A'B'$ (not shown in the figure), which is moving with velocity $V$
along the positive direction of the axis $x\equiv x'$. The train is coming from the negative values of $x$. $A$ and
$B$ are the extreme points of the station's platform, $O$ its center, and $L_0$ its proper length. The clocks of $O$, $A$,
and $B$ have been previously synchronized. At the time $t=0$, $A$ and $B$ send two flashes towards $O$. We assume that
at $t=0$, $O'$ meets $O$, and the clock of $O'$ is set at $t'=0$. In Einstein's book, the two light flashes consist of two light strokes.
The colors \textcolor[rgb]{1.00,0.00,0.00}{Red} e \textcolor[rgb]{0.00,0.00,1.00}{Blue} have only a graphical meaning.}}\label{re-sim-ein}
\end{figure}

In the platform's IRS (Fig. \ref{re-sim-ein}), the two flashes reach the center of the platform $O$ at the same time $t_O=L_0/2c$, where $L_0$ is the proper
length of the platform. Therefore, the  duration of the phenomenon whose initial event is
`the \textcolor[rgb]{1.00,0.00,0.00}{Red} flash reaches $O$' and its final event is `the \textcolor[rgb]{0.00,0.00,1.00}{Blue} flash reaches
$O$' is null. Hence, the duration of the same phenomenon in $O'$ will be $\Gamma \times 0=0$, i.e.,  also for $O'$, the two flashes
simultaneously reach the center of the station's platform.
Therefore:
\begin{equation}\label{pertantomg}
t'_{AO}=t'_{BO}=t'_O,
\end{equation}
where $t'_{AO}$ is the time at which the \textcolor[rgb]{1.00,0.00,0.00}{Red} flash ($A \rightarrow O$) reaches $O$, and $t'_{BO}$ is
the time at which the \textcolor[rgb]{0.00,0.00,1.00}{Blue} flash ($B\rightarrow O$) reaches $O$. We shall use this result later.
\par\noindent
\textsf{Teachers should stress that we have used the time dilation formula in the IRS of the `moving' train, thus reaffirming that
time dilation is a symmetrical phenomenon. As noticed in the main text (point \ref{TD}, on p. \pageref{TD}), students tend to believe it is not so.
}
\par
According to $O'$, the \textcolor[rgb]{1.00,0.00,0.00}{Red} flash reaches the center of the platform $O$ at the time:
\begin{equation}\label{a-o}
t'_O=t'_{A}+\left[\frac{L_0}{2\Gamma} -(t'_O-t'_A)V \right]\frac{1}{c}.
\end{equation}
Solving for $( t'_O-t'_{A})$, we get:
\begin{equation}\label{t'ao}
t'_O-t'_{A}=   \frac{L_0}{2\Gamma c}\frac{1}{1+B}.
\end{equation}
Similarly, for the \textcolor[rgb]{0.00,0.00,1.00}{Blue} flash:
\begin{equation}\label{t'bo}
t'_O-t'_{B}=   \frac{L_0}{2\Gamma c}\frac{1}{1-B}.
\end{equation}
Subtracting member by member Eq. (\ref{t'ao}) from Eq. (\ref{t'bo}):
\begin{equation}\label{diffe-emiss}
t'_{A}-t'_{B}=\frac{L_0}{2\Gamma c}\left( \frac{1}{1-B} -\frac{1}{1+B} \right)= \frac{L_0}{c} \Gamma B.
\end{equation}
So, we have proved that, in the train's reference frame, the emissions of the two flashes are not simultaneous.
\par
Summing member by member the same equations we obtain:
\begin{equation}\label{sommadueultime}
t'_O=\frac{L_0}{2\Gamma c}+\frac{t'_{A}+t'_{B}}{2}.
\end{equation}
For the homogeneity of the variable $t$, if  $t'_{A}$ and  $t'_{B}$ satisfy Eq. (\ref{diffe-emiss}),
also the pair $t''+t'_{A}$ and  $t''+t'_{B}$, $t''$ being an arbitrary value, satisfy the same equation. Since the value
of $t'_O$ can not depend on the pair chosen, it follows that the sum $t'_{A}+t'_{B}$ of Eq. (\ref{sommadueultime}) must be zero, namely,
the equality $t'_{A}= -t'_{B}$ must hold. Therefore:
\begin{equation}\label{ecco!}
t'_O=\frac{L_0}{2\Gamma c}.
\end{equation}
Moreover, from Eq. (\ref{diffe-emiss}) we obtain:
\begin{eqnarray}
t'_A&=& \frac{L_0}{2 c}\Gamma B \label{t'a}\\
t'_B &=& - \frac{L_0}{2 c}\Gamma B.\label{t'b}
\end{eqnarray}
Now, we can easily calculate
the values of $x'_O(t'_O)$ and  $x'_{A}(t'_A) $.
It turns out that:
\begin{equation}\label{x'o}
x'_O(t'_O) = -V(t'_O-0)=-V t'_O=-\frac{L_0}{2}B\Gamma.
\end{equation}
The value of $x'_A$ obeys the equation:
\begin{equation}\label{x'a}
x'_O(t'_O)-x'_{A}(t'_A) =  c(t'_O-t'_A).
\end{equation}
Using Eqs. (\ref{x'o}) and (\ref{t'a}), we get:
\begin{equation}\label{x'A}
x'_A=-\frac{L_0}{2}\Gamma.
\end{equation}
Similarly, we obtain:
\begin{equation}\label{x'B}
x'_B=\frac{L_0}{2}\Gamma.
\end{equation}
\textsf{As the reader might have noticed, differently from section \ref{train}, we have calculated first the times of the events and,
subsequently, the $x$ coordinates. Teachers should choose between the two approaches or use them both.
}
\section{Gravitational red shift with special relativity}\label{redshift}
Consider an excited atom at rest
in the Earth's gravitational field, assumed to be spherically symmetric (see, \cite[276-278]{erq}). The rest energy
of the pair (atom + Earth) will be:
\begin{equation}\label{atomo+terra}
{\mathcal E}_{0}(r) =  M c^2+ \left( mc^2+ \Delta E_\infty\right)
- {{GM}\over {r}}{ {\left( m+ {{\Delta E_\infty}\over {c^2}}\right) } },
\end{equation}
where $M$ and $m$ are the Earth's and atom's mass, respectively, $G$ the gravitational constant, $r$ the distance of the atom from the Earth center, and $\Delta E_\infty$ the atom's excitation energy in the absence of the gravitational field.
Then, the available energy for the atom's emission will be:
\begin{equation}\label{fotogr}
\Delta E (r) = \Delta E_\infty \left( 1 - {{GM}\over {r}}{ { {{1}\over {c^2}} }  } \right) = \Delta E_\infty  \left( 1 + {{\chi(r)}\over {c^2}} \right),
\end{equation}
$\chi$ being the Earth's Newtonian gravitational potential at the point $r$.
If we take two points $R$ and $R+h$ ($R$ being the Earth's radius), we have:
\begin{eqnarray}
\Delta E(R)&=&\Delta E_\infty  \left(1-\frac{GM}{c^2}\frac{1}{R}
\right)\label{terraR}\\
&&\nonumber\\
\Delta E(R+h)&=& \Delta E_\infty  \left(1-\frac{GM}{c^2}\frac{1}{R+h} \right). \label{terrah}
\end{eqnarray}
If $h\ll R$, then:
$$\frac{1}{R+h}\approx \frac{1}{R}\left(1-\frac{h}{R}\right)$$
Therefore:
\begin{equation}\label{Terra} \Delta E(R+h)\approx \Delta E(R) + \Delta E_\infty{{gh}\over {c^2}}\approx \Delta E(R)\left(1 + {{gh}\over {c^2}} \right),
\end{equation}
where  $g$ is the Earth's gravitational field.
In 1960, Pound and Rebka corroborated Eq. (\ref{Terra}) in a famous experiment using photons emitted or absorbed by nuclei without recoil  (M\"{o}ssbauer effect) \cite{pound}. In Pound and Rebka's experiment, $h=22.55 {\rm m}$ and $gh/c^2=2.46\times 10^{-15}$.
In the last passage of Eq. (\ref{Terra}), we have put $\Delta E(R)=\Delta E_\infty$. Indeed, for $R=6378 \times 10^3{\rm m}$ and $g=9.8 {\rm ms^{-2}}$, we have $\Delta E (r)=\Delta E_\infty(1-6.95\times 10^{-10})$ and $gh/c^2=2.46\times 10^{-15}$. Hence, if in Eq. (\ref{Terra}), we put $\Delta E(R)$ instead of $\Delta E_\infty$ we introduce an error of the order of $10^{-10}\times 10^{-15}=10^{-25}$ which is $10^{10}$ smaller than the value we must measure.
\par
We can re-write Eq. (\ref{Terra}) in terms of frequencies:
\begin{equation}\label{terranu}
\nu (R+h) \approx
\nu  (R)\left(1 + {{gh}\over {c^2}} \right),
\end{equation}
and the above equation in terms of periods:
\begin{equation}\label{periodi}
T (R+h) \approx T  (R)\left(1 - {{gh}\over {c^2}} \right).
\end{equation}
This equation is a particular form of a more general one:
\begin{equation}\label{ppgen}
T(r)\approx T_\infty \left(1-\frac{\chi(r)}{c^2} \right); \chi<0,
\end{equation}
that is valid for constant and small gravitational fields. Since atomic clocks use as a frequency standard a frequency associated with a photon emitted or absorbed by an atom, Eq. (\ref{ppgen}) shows that a gravitational field modifies the fundamental period of atomic clocks. Precisely, the fundamental period of an atomic clock is increased by a gravitational field (through its gravitational potential).
\end{appendices}

\vskip4mm\par

\end{document}